\documentclass{aa}  
\usepackage{xcolor}
\usepackage{graphicx} 
\usepackage{physics}
\usepackage{amsmath}
\usepackage{amssymb}

\definecolor{lightgray}{rgb}{0.83, 0.83, 0.83}
\definecolor{lightblue}{rgb}{0.67, 0.84, 0.90}
\definecolor{lightgreen}{rgb}{0.56, 0.93, 0.56}

\usepackage{natbib}
\bibpunct{(}{)}{;}{a}{}{,} 

\usepackage{threeparttable}

\usepackage[varg]{txfonts}
\usepackage{hyperref}
\usepackage{multirow}
\usepackage{color}
\definecolor{green}{rgb}{0.3,0.7,0.}
\definecolor{purple}{rgb}{0.77, 0.29, 0.55}

\hypersetup{
    colorlinks=true,       
    linkcolor=blue,          
    citecolor=blue,        
    filecolor=blue,      
    urlcolor=blue           
}

\begin{document}

\title{The Evolution of Pop III.1 Protostars Powered by Dark Matter Annihilation.
II. 
Dependence on WIMP Properties
}

\titlerunning{Pop III.1 Protostars - Dependence on WIMP Properties}
\author{Konstantinos Topalakis\inst{1}, Devesh Nandal\inst{2}, Jonathan C. Tan\inst{2,3}}
\authorrunning{Topalakis et al.}

\institute{Department of Physics, Gothenburg University, 412 96 Gothenburg, Sweden \and Department of Astronomy and Virginia Institute for Theoretical Astronomy, University of Virginia, 530 McCormick Rd, Charlottesville, VA 22904, USA \and Dept. of Space, Earth \& Environment, Chalmers University of Technology, Chalmersgatan 4, 412 96 Gothenburg, Sweden}

\date{}

\abstract{
The rapid appearance of supermassive black holes (SMBHs) at $z\gtrsim7$ requires efficient pathways to form massive black hole seeds. We investigate whether annihilation of weakly interacting massive particles (WIMPs) can alter primordial (Pop III.1) protostellar evolution sufficiently to enable formation of such ``heavy'' seeds. Using the one-dimensional Geneva stellar-evolution code (GENEC) with an implemented Gould single-scatter capture module, we compute a grid of protostellar evolution models covering ambient WIMP mass densities $\rho_\chi=10^{12}$–$10^{16}\ \mathrm{GeV\,cm^{-3}}$, WIMP masses $m_\chi=30$–$3000\ \mathrm{GeV}$, spin-dependent cross sections $\sigma_{\rm SD}=10^{-42}$–$10^{-40}\ \mathrm{cm^2}$, and baryonic accretion rates $\dot{M_*}=(1-3)\times10^{-3}\, M_\odot \,{\rm yr}^{-1}$.
We find a robust bifurcation of outcomes. For sufficiently high ambient dark matter density ($\rho_\chi\gtrsim5\times10^{14}\ \mathrm{GeV\,cm^{-3}}$) and capture efficiency ($\sigma_{\rm SD}\gtrsim10^{-41}\ \mathrm{cm^2}$) WIMP annihilation supplies enough energy to inflate protostars onto extended, cool (Hayashi-track) configurations that dramatically suppress ionizing feedback and permit uninterrupted growth to $\sim10^{5}\,M_\odot$. Lighter WIMPs and larger $\sigma_{\rm SD}$ favour earlier and stronger annihilation support; heavier WIMPs delay the effect. For our fiducial case, WIMP masses $<$3 TeV are essential for allowing growth to the supermassive regime, otherwise the protostar evolves to the compact, feedback-limited regime that results in ``light'' seeds. These results indicate that, under plausible halo conditions, DM annihilation provides a viable channel for forming heavy black hole seeds.}


\maketitle

\section{Introduction}

The rapid appearance of supermassive black holes (SMBHs), including some billion‐solar‐mass quasars, at redshifts $z\gtrsim7$ continues to challenge models of early black hole seed formation, demanding efficient pathways from pristine gas to SMBHs within a few hundred million years after the Big Bang \citep{Fan2003,Mortlock2011,Wang2021,Yang2021,2023ApJ...959...39H,2024A&A...691A.145M}. In this context, Population III.1 stars, i.e., metal‐free protostars born in locally isolated first-forming dark‐matter minihalos, i.e., that have not been subject to any significant external feedback, have emerged as a promising channel for producing abundant heavy seeds \citep{Banik2019,Singh2023,2025MNRAS.536..851C,2025MNRAS.542.1532S,2025arXiv250702058S,2025ApJ...989L..47T} \citep[for a review see][]{Tan2024}. In particular, in the Pop III.1 scenario, WIMP annihilation heating is crucial for its impact on protostellar collapse \citep{Spolyar2008,Natarajan2009} and evolution \citep{Freese2010,Ilie2012,RindlerDaller2015,Ilie2021}. The general trend is to dramatically inflate the protostar, suppress ionizing feedback that would be present if the protostar was accreting on the main sequence \citep{2004ApJ...603..383T,McKeeTan2008}, and thus potentially permit uninterrupted growth to $\sim10^5\,M_\odot$, with this characteristic mass scale set by the baryonic content of the minihalo. 

Key features of the Pop III.1 seeding model are that it can explain both the overall cosmic abundance of SMBHs, i.e., $n_{\rm SMBH}\gtrsim10^{-2}\:{\rm cMpc}^{-3}$ \citep{2025arXiv250117675C}, and also why there is an apparent dearth of intermediate-mass black holes (IMBHs) and/or a break in the SMBH mass function near $\sim 10^5\:M_\odot$ \citep{2025MNRAS.541..429M}. We note that other mechanisms for forming SMBHs struggle to reproduce these features. For example, the ``direct collapse'' mechanism \citep[e.g.,][]{2003ApJ...596...34B} involving high accretion rates from massive ($\sim10^8\:M_\odot$), FUV-irradiated, atomically-cooled halos struggles to match the cosmic abundance of SMBHs by several orders of magnitude \citep[e.g.,][]{2016ApJ...832..134C,2019Natur.566...85W,2025OJAp....8E..88O}. Models of SMBH formation from dense star clusters have been proposed \citep[e.g.,][]{2004ApJ...604..632G,2006MNRAS.368..141F,2022MNRAS.512.6192S}, however the required conditions, even to make IMBHs, are almost never seen in local star-forming regions \citep[e.g.,][]{2014prpl.conf..149T}. Furthermore, these models would predict much larger abundances of IMBHs compared to SMBHs, while, as mentioned, there is very limited evidence for the existence of such IMBHs \citep[e.g.,][]{2017MNRAS.464.2174B,2020ARA&A..58..257G,2025MNRAS.541..429M}. Super-Eddington growth from ``light'' seed black holes has been proposed as an alternative method of forming the high-redshift SMBH population \citep[e.g.,][]{2024A&A...686A.256L}, however, there is limited evidence for such super-Eddington growth rates in observed quasar populations \citep[e.g.,][]{Yang2021,2022ApJS..263...42W,2024ApJ...966...85Z}. Furthermore, the low metallicity recently reported around a high redshift SMBH by \citet{2025arXiv250522567M} appears to be difficult to reconcile with direct collapse, dense star cluster and super-Eddington growth scenarios.

In Paper I of this series, \citet{Nandal2025} implemented the \citet{Gould1987} single‐scatter capture formalism into the GENEC stellar‐evolution code and explored a grid of nine models spanning ambient WIMP densities $10^{12}$–$10^{16}$ GeV cm$^{-3}$ and accretion rates $10^{-3}$–$10^{-1}\,M_\odot\,$yr$^{-1}$.  They demonstrated a critical density threshold ($\rho_\chi\gtrsim5\times10^{14}$ GeV cm$^{-3}$) for supermassive growth at a fiducial $\dot M=3\times10^{-3}\,M_\odot\,$yr$^{-1}$, and showed that enhanced WIMP heating not only inflates the stellar envelope—quenching ionizing photon output by orders of magnitude—but also postpones general‐relativistic collapse to masses $\gtrsim10^6\,M_\odot$.

While the Paper I study fixed the WIMP mass ($m_\chi=100\,$GeV) and adopted benchmark scattering cross‐sections consistent with recent LZ limits, the broader dark‐matter microphysics remains highly uncertain. Laboratory and astrophysical searches constrain spin‐dependent and spin‐independent cross sections only down to $\sigma\sim10^{-41}$–$10^{-47}\,$cm$^2$ \citep{LZ2024}, and WIMP masses from a few tens of GeV up to multi‐TeV remain viable. Changes in $m_\chi$ and $\sigma$ alter both the capture efficiency (via kinematic suppression for $m_\chi\gg m_N$) and the equilibrium timescale for annihilation heating \citep{Freese2009,Ilie2012}.

In this second paper, we extend the parameter survey of Pop III.1 evolution by exploring a three‐dimensional grid  for the WIMPs parameters in
\begin{align}
\rho_\chi &= 10^{12},\,10^{13},\,10^{14},\,10^{15},\,10^{16}\;\mathrm{GeV\,cm^{-3}}, \nonumber \\
m_\chi &= 30,\;50,\;100,\;300,\;1000,\;3000\;\mathrm{GeV}, \nonumber \\
\sigma_{\rm SD} &= 10^{-40},\,10^{-41},\,10^{-42}\;\mathrm{cm^2}. \nonumber
\end{align}
All other WIMP parameters (thermal annihilation cross section $\langle\sigma v\rangle=3\times10^{-26}$ cm$^3$s$^{-1}$, spin‐independent cross section $\sigma_{\rm SI}=10^{-47}$ cm$^2$, halo velocity dispersion $v_\chi=10$ km s$^{-1}$) and the stellar accretion physics follow the fiducial setup of \citet{Nandal2025}.

Our goals are threefold:\\
1. Quantify how variations in $m_\chi$ and $\sigma_{\rm SD}$ shift the critical $\rho_\chi$ needed for supermassive growth and feedback suppression.\\
2. Assess the sensitivity of protostellar structure, ionizing‐photon output, and Kelvin–Helmholtz contraction phases to dark‐matter microphysics.\\
3. Identify the parameter regimes where dark matter heating and feedback lead to the formation of massive black hole seeds ($\sim10^5\,M_\odot$), and delay or suppress the collapse of protostars in the early universe.

By systematically probing beyond the “typical” WIMP mass and cross‐section values, this study delivers a robust understanding of how uncertainties in dark‐matter properties translate into the landscape of primordial supermassive star formation and early black‐hole seeding.

\section{Methods}\label{Sec:Methods}



In this work we adopt the same physical and numerical framework introduced in \citet{Nandal2025}, namely the single‐scatter WIMP capture formalism of \citet{Gould1987} coupled to the one–dimensional GENEC stellar evolution code \citep{Eggenberger2008, Nandal2024a}. Below we summarize the key ingredients and highlight only the extensions needed for our enlarged parameter grid.

\subsection{Stellar Evolution with GENEC}
\label{sec:genec}

GENEC solves the four stellar structure equations (mass, hydrostatic equilibrium, energy transport, and energy conservation) via the Henyey method \citep{Eggenberger2008}. At each timestep we pass the converged profiles of density $\rho(r)$, temperature $T(r)$, and escape velocity $v_{\rm esc}(r)$ to the DM module, integrate the DM capture rate $C_{\rm c}(t)$, update the DM particles number $N_\chi$ via the analytic solution, and then add the DM annihilation energy $\varepsilon_\chi(r)$ to the local energy generation. Pre‐main‐sequence growth is modelled by imposing a constant gas accretion rate $\dot M_*$, inserting new outer shells of pristine composition, and reconverging the structure before computing capture. This constant accretion of baryonic matter is implemented in \citet{Nandal2024c}. Radiative feedback is treated as in Paper I of our series \citep{Nandal2025}, with accretion capped by the photo‐evaporation mass loss following \citet{McKeeTan2008}.

All models considered in this study have been tested for gravitational stability based on the criteria outlined in \citep{Baumgarte1999,Haemmerle2021,Nandal2024c}. This evaluation ensures that the conditions for the formation and evolution of Pop III.1 stars are properly accounted for, particularly in terms of their potential to avoid or delay collapse due to dark matter heating and feedback effects.

\subsection{Capture Rate}
\label{sec:capture}

Starting from the differential form of the DM capture rate \citep[explained in detail in][]{Nandal2025}, 
\begin{equation}
    \frac{dC_{\rm c}}{dm} =
    \sqrt{\frac{6}{\pi}}\, 
    \frac{\sigma_{\rm eff}\,\rho_\chi}{m_\chi}\,
    \frac{v_{\rm esc}^{2}}{v_\chi}\,
    \frac{\mathcal{P}}{2\sqrt{3/2}\,A^{2}}\,,
\label{eq:dcdm}
\end{equation}
where $\sigma_{\rm eff}$ is the scattering cross-section of the DM particles, $m_\chi$ the WIMP mass, $v_\chi$ the one-dimensional halo velocity dispersion, and $\mathcal{P}$ arises from performing the angular integration over the Maxwellian halo velocity distribution.

Substituting the kinematic factor $A^{2} = \frac{3\,v_{\rm esc}^{2}\,\mu}{2\,v_\chi^{2}\,\mu_{\rm red}^{2}}$ where $A^2\gg1$:
\begin{equation}
    [\mathcal{P}]\propto A_+A_- \propto A^2\propto v_{\rm esc}^2.
\end{equation}
All the other terms in Eq.~\ref{eq:dcdm} are constants or reflect halo properties. Thus, integrating for the total stellar mass $M_*$:
\begin{equation}
    \frac{dC_{c}}{dm}\propto v_{\rm esc}^2(r) \quad\Longrightarrow\quad C_{c} \propto \int_0^{M_*}\!v_{\rm esc}^2\,dm\,.
\end{equation}
Therefore, we recover the scaling that contains both the stellar mass $M_*$ and radius $R_*$ for the number capture rate
\begin{equation}
    C_{c} \propto \frac{1}{m_\chi} \frac{M_*^2}{R_*}\,.
\label{eq:capture}
\end{equation}
or equivalently for the mass capture rate
\begin{equation}
    m_\chi C_{c} \propto \frac{M_*^2}{R_*}\,.
\label{eq:capture2}
\end{equation}

Capture in our implementation is only counted when the energy transfer in a single elastic scatter is sufficient to make the WIMP bound to the star (i.e. its post-scatter speed is below the local escape speed at the scattering radius). We use the standard Gould single-scatter formalism. The code integrates the halo velocity distribution and the per-shell kinematics to compute the probability that a passing WIMP, upon scattering with a local nucleus, will lose enough energy to be gravitationally captured.

\subsection{Total WIMP mass in equilibrium}

From the equilibrium solution of the total number of WIMPs, $N_\chi(t)$, bound in the protostar via
\begin{equation}
    \frac{dN_\chi}{dt} \;=\; C_{c} \;-\; A\,N_\chi^2,
\label{eq:dNdt}
\end{equation}
we have
\[
N_{\chi,\infty} = \sqrt{\frac{C_c}{A}}\,,\qquad A = \frac{\langle\sigma_a v\rangle}{V_{\rm eff}}\,,\quad V_{\rm eff} \approx \sqrt2\,\pi^{-3/2}\,r_\chi^3\,.
\]
Thus the total dark‐matter mass, $M_{\rm \chi, tot}$, bound in the star is
\begin{equation}
    \begin{split}
        M_{\rm \chi, tot} &= m_\chi \, N_{\chi,\infty} = m_\chi \,\sqrt{\frac{C_c}{A}} \\
        &= m_\chi \,\sqrt{\frac{C_c}{\langle\sigma_a v\rangle}} \left(\sqrt{2}\,\pi^{-3/2}\,r_\chi^3 \right)^{1/2} \\
        &= m_\chi \,\sqrt{\frac{C_c}{\langle\sigma_a v\rangle}} \left(2^{1/4}\,\pi^{-3/4}\,r_\chi^{3/2} \right) \\
        &\propto r_\chi^{3/2} \, m_\chi \, \sqrt{C_c}.
    \end{split}
\end{equation}
Recall the scale radius where the WIMPs are concentrated in the star
\begin{equation}
    r_\chi = \sqrt{\frac{3k_B T_c}{2\pi G \rho_c\,m_\chi}}  \quad\Longrightarrow\quad r_\chi^{3/2} \propto \left(\frac{T_c}{\rho_c}\right)^{3/4} m_\chi^{-3/4}\,.
\end{equation}
Substituting into $M_{\rm \chi, tot}$ gives
\begin{equation}
    M_{\rm \chi, tot} \propto \left(\frac{T_c}{\rho_c}\right)^{3/4} \, m_\chi^{1-3/4} \, \sqrt{C_c} = \left(\frac{T_c}{\rho_c}\right)^{3/4} \, m_\chi^{1/4} \, \sqrt{C_c}.
\end{equation}
Since $C_c \propto m_\chi^{-1}$ (from Eq.~\ref{eq:dcdm}), we find
\begin{equation}
    M_{\rm \chi, tot} \propto \left(\frac{T_c}{\rho_c}\right)^{3/4} \, m_\chi^{1/4} \, m_\chi^{-1/2} = \left(\frac{T_c}{\rho_c}\right)^{3/4} \, m_\chi^{-1/4}.
\label{eq:Mchi}
\end{equation}

\subsection{Parametric Study and Initial Parameters}
\label{sec:perameters}

We perform a $5 \times 6 \times 3 \times 2$ grid of models, varying:
\begin{itemize}
    \item Ambient WIMP density: $\rho_\chi = 10^{12},\,10^{13},\,10^{14},\,10^{15},\,10^{16}$ GeV cm$^{-3}$,
    \item WIMP mass: $m_\chi = 30,\,50,\,100,\,300,\,1000,\,3000$ GeV,
    \item Spin‐dependent scattering cross section: $\sigma_{\rm SD} = 10^{-42},\,10^{-41},\,10^{-40}$ cm$^2$,
    \item Baryonic accretion rate: $\dot{M}_* = 10^{-3},\,3\times10^{-3}\;M_\odot\,\mathrm{yr}^{-1}$,
\end{itemize}
while holding fixed the annihilation cross section $\langle\sigma_a v\rangle=3\times10^{-26}$ cm$^3$s$^{-1}$, spin‐independent scattering cross section (which has only a minor influence) $\sigma_{\rm SI}=10^{-47}$ cm$^2$, and halo velocity dispersion $v_\chi=10$ km s$^{-1}$. Each simulation begins from a 2 $M_\odot$ protostellar seed at an age of 9 yr, with H and He mass fractions of $X=0.7516$ and $Y=0.2484$, and is evolved until either radiative feedback halts accretion or numerical termination at $\sim10^5\,M_\odot$.

By systematically varying $m_\chi$ and $\sigma_{\rm SD}$, in addition to $\rho_\chi$, we quantify how dark‐matter microphysics influences protostellar inflation and ionizing‐photon output.

\begin{table*}[h]
    \centering
    \caption{Initial parameters of the models ($\dot{M}_*$, $\rho_{\chi}$, $m_\chi$, $\sigma_{SD}$) and the final values for stellar mass ($M_{*f}$), age $t_{*f}$, Eddington factor $\Gamma_{\rm Edd}$, convective core mass fraction $M_{\rm cc}$, final hydrogen central abundance $X_{^1 \rm H}$, and WIMP quantities. The columns are as follows: the first four columns represent the initial mass accretion rate ($\dot{M}_*$) in $[M_\odot\, \mathrm{yr^{-1}}]$, the initial background WIMP density ($\rho_{\chi}$) in $[GeV\, cm^{-3}]$, the mass of each WIMP ($m_\chi$) in $GeV/c^2$ and the spin-dependent scattering cross-section $\sigma_{SD}$ in $cm^2$, followed by the final stellar mass $M_{*f}$ in $[M_\odot]$, final stellar age $t_{*f}$ in years, Eddington factor $\Gamma_{\rm Edd}$, final convective core mass fraction $M_{\rm cc}$, final hydrogen central abundance $X_{^1 \rm H}$, initial WIMP number $N_{\chi,i}$, initial WIMP mass $M_{\chi,i}$ in $[M_\odot]$, final WIMP number $N_{\chi,f}$, and final WIMP mass $M_{\chi,f}$ in $[M_\odot]$.}
    \resizebox{2\columnwidth}{!}{
    \begin{tabular}{|cccc|ccccccccc|}
        \hline\hline
        $\dot{M}_*$ [$M_\odot$\,yr$^{-1}$] & $\rho_{\chi}$ [GeV\,cm$^{-3}$] & $m_{\chi}$ [GeV] & $\sigma_{SD}$ [cm$^2$] & $M_{*f}$ [M$_\odot$] & $t_{*f}$ [yr] & $\Gamma_{\rm Edd}$ & $M_{\rm cc}$ & $X_{^1\mathrm{H}}$ & $N_{\chi,i}$ & $M_{\chi,i}$ [M$_\odot$] & $N_{\chi,f}$ & $M_{\chi,f}$ [M$_\odot$]\\

        $10^{-3}$ & 0 & 0 & 0 & 243 & $2.409 \cdot 10^{5}$ & 0.5673 & 0.8880 & 0.7380 & 0 & 0 & 0 & 0 \\ 
        $10^{-3}$ & $10^{12}$ & 100 & $10^{-41}$ & 246 & $2.440 \cdot 10^{5}$ & 0.5691 & 0.8901 & 0.7392 & $1.713 \cdot 10^{47}$ & $1.535 \cdot 10^{-8}$ & $5.527 \cdot 10^{49}$ & $4.954 \cdot 10^{-6}$\\ 
        $10^{-3}$ & $10^{13}$ & 30 & $10^{-41}$ & 249.5 & $2.475 \cdot 10^{5}$ & 0.5669 & 0.9017 & 0.7492 & $1.489 \cdot 10^{48}$ & $4.004 \cdot 10^{-8}$ & $4.939 \cdot 10^{50}$ & $1.328 \cdot 10^{-5}$\\ 
        $10^{-3}$ & $10^{13}$ & 50 & $10^{-41}$ & 248 & $2.461 \cdot 10^{5}$ & 0.5657 & 0.9015 & 0.7490 & $7.777 \cdot 10^{47}$ & $3.485 \cdot 10^{-8}$ & $2.680 \cdot 10^{50}$ & $1.201 \cdot 10^{-5}$\\ 
        $10^{-3}$ & $10^{13}$ & 100 & $10^{-41}$ & 249.5 & $2.475 \cdot 10^{5}$ & 0.5671 & 0.9011 & 0.7493 & $5.416 \cdot 10^{47}$ & $4.854 \cdot 10^{-8}$ & $1.787 \cdot 10^{50}$ & $1.602 \cdot 10^{-5}$ \\ 
        $10^{-3}$ & $10^{13}$ & 300 & $10^{-41}$ & 247 & $2.451 \cdot 10^{5}$ & 0.5648 & 0.9024 & 0.7495 & $6.888 \cdot 10^{46}$ & $1.852 \cdot 10^{-8}$ & $2.240 \cdot 10^{49}$ & $6.023 \cdot 10^{-6}$\\ 
        $10^{-3}$ & $10^{13}$ & 1000 & $10^{-41}$ & 250 & $2.480 \cdot 10^{5}$ & 0.5684 & 0.8986 & 0.7462 & $9.498 \cdot 10^{45}$ & $8.513 \cdot 10^{-9}$ & $3.134 \cdot 10^{49}$ & $2.809 \cdot 10^{-5}$\\ 
        $10^{-3}$ & $10^{14}$ & 100 & $10^{-41}$ & 348.4 & $3.464 \cdot 10^{5}$ & 0.5934 & 0.9461 & 0.7516 & $1.713 \cdot 10^{48}$ & $1.535 \cdot 10^{-7}$ & $9.401 \cdot 10^{50}$ & $8.426 \cdot 10^{-5}$\\
        $10^{-3}$ & $10^{15}$ & 30 & $10^{-41}$ & 136714 & $1.367 \cdot 10^{8}$ & 1.3624 & 1.0000 & 0.7516 & $1.489 \cdot 10^{49}$ & $4.004 \cdot 10^{-7}$ & $6.672 \cdot 10^{55}$ & 1.794\\ 
        $10^{-3}$ & $10^{15}$ & 50 & $10^{-41}$ & 139066 & $1.391 \cdot 10^{8}$ & 1.3877 & 1.0000 & 0.7516 & $7.777 \cdot 10^{48}$ & $3.485 \cdot 10^{-7}$ & $3.597 \cdot 10^{55}$ & 1.612\\ 
        $10^{-3}$ & $10^{15}$ & 100 & $10^{-42}$ & 350.7 & $3.487 \cdot 10^{5}$ & 0.5947 & 0.9453 & 0.7516 & $1.713 \cdot 10^{48}$ & $1.535 \cdot 10^{-7}$ & $9.498 \cdot 10^{50}$ & $8.513 \cdot 10^{-5}$\\ 
        $10^{-3}$ & $10^{15}$ & 100 & $10^{-41}$ & 139164 & $1.392 \cdot 10^{8}$ & 1.3705 & 1.0000 & 0.7516 & $5.416 \cdot 10^{48}$ & $4.854 \cdot 10^{-7}$ & $2.073 \cdot 10^{55}$ & 1.858\\ 
        $10^{-3}$ & $10^{15}$ & 100 & $10^{-40}$ & 55970 & $5.597 \cdot 10^{7}$ & 4.2459 & 1.0000 & 0.7516 & $1.713 \cdot 10^{49}$ & $1.535 \cdot 10^{-6}$ & $4.890 \cdot 10^{55}$ & 4.383 \\
        $10^{-3}$ & $10^{15}$ & 300 & $10^{-41}$ & 139027 & $1.390 \cdot 10^{8}$ & 1.3891 & 1.0000 & 0.7516 & $6.888 \cdot 10^{47}$ & $1.852 \cdot 10^{-7}$ & $3.898 \cdot 10^{54}$ & 1.048\\ 
        $10^{-3}$ & $10^{15}$ & 1000 & $10^{-41}$ & 134118 & $1.341 \cdot 10^{8}$ & 1.3756 & 1.0000 & 0.7516 & $9.498 \cdot 10^{46}$ & $8.513 \cdot 10^{-8}$ & $1.922 \cdot 10^{54}$ & 1.723\\ 
        $10^{-3}$ & $10^{16}$ & 100 & $10^{-41}$ & 50197 & $5.020 \cdot 10^{7}$ & 4.1679 & 1.0000 & 0.7516 & $1.713 \cdot 10^{49}$ & $1.535 \cdot 10^{-6}$ & $4.350 \cdot 10^{55}$ & 3.899\\     

        $3\cdot10^{-3}$ & 0 & 0 & 0 & 436.3 & $1.448 \cdot 10^{5}$ & 0.6887 & 0.9197 & 0.7342 & 0 & 0 & 0 & 0 \\ 
        $3\cdot10^{-3}$ & $10^{12}$ & 100 & $10^{-41}$ & 442.7 & $1.469 \cdot 10^{5}$ & 0.6920 & 0.9175 & 0.7348 & $1.713 \cdot 10^{47}$ & $1.535 \cdot 10^{-8}$ & $9.954 \cdot 10^{49}$ & $8.921 \cdot 10^{-6}$\\ 
        $3\cdot10^{-3}$ & $10^{13}$ & 30 & $10^{-41}$ & 437.1 & $1.450 \cdot 10^{5}$ & 0.6832 & 0.9317 & 0.7446 & $1.489 \cdot 10^{48}$ & $4.004 \cdot 10^{-8}$ & $8.632 \cdot 10^{50}$ & $2.321 \cdot 10^{-5}$\\ 
        $3\cdot10^{-3}$ & $10^{13}$ & 50 & $10^{-41}$ & 437.6 & $1.452 \cdot 10^{5}$ & 0.6840 & 0.9280 & 0.7446 & $7.777 \cdot 10^{47}$ & $3.485 \cdot 10^{-8}$ & $4.731 \cdot 10^{50}$ & $2.120 \cdot 10^{-5}$\\ 
        $3\cdot10^{-3}$ & $10^{13}$ & 100 & $10^{-41}$ & 438 & $1.453 \cdot 10^{5}$ & 0.6837 & 0.9212 & 0.7446 & $5.416 \cdot 10^{47}$ & $4.854 \cdot 10^{-8}$ & $3.139 \cdot 10^{50}$ & $2.813 \cdot 10^{-5}$\\ 
        $3\cdot10^{-3}$ & $10^{13}$ & 300 & $10^{-41}$ & 439.9 & $1.460 \cdot 10^{5}$ & 0.6869 & 0.9209 & 0.7448 & $6.888 \cdot 10^{46}$ & $1.852 \cdot 10^{-8}$ & $3.991 \cdot 10^{49}$ & $1.073 \cdot 10^{-5}$\\ 
        $3\cdot10^{-3}$ & $10^{13}$ & 1000 & $10^{-41}$ & 439.1 & $1.457 \cdot 10^{5}$ & 0.6822 & 0.9248 & 0.7424 & $9.498 \cdot 10^{45}$ & $8.513 \cdot 10^{-9}$ & $5.504 \cdot 10^{49}$ & $4.933 \cdot 10^{-5}$\\ 
        $3\cdot10^{-3}$ & $10^{14}$ & 100 & $10^{-41}$ & 702.1 & $2.334 \cdot 10^{5}$ & 0.7088 & 0.9708 & 0.7516 & $1.713 \cdot 10^{48}$ & $1.535 \cdot 10^{-7}$ & $1.895 \cdot 10^{51}$ & $1.698 \cdot 10^{-4}$\\ 
        $3\cdot10^{-3}$ & $10^{15}$ & 30 & $10^{-41}$ & 138571 & $4.619 \cdot 10^{7}$ & 1.3838 & 1.0000 & 0.7516 & $1.489 \cdot 10^{49}$ & $4.004 \cdot 10^{-7}$ & $6.721 \cdot 10^{55}$ & 1.807 \\ 
        $3\cdot10^{-3}$ & $10^{15}$ & 50 & $10^{-41}$ & 137256 & $4.575 \cdot 10^{7}$ & 1.3644 & 1.0000 & 0.7516 & $7.777 \cdot 10^{48}$ & $3.485 \cdot 10^{-7}$ & $3.535 \cdot 10^{55}$ & 1.584 \\ 
        $3 \cdot 10^{-3}$ & $10^{15}$ & 100 & $10^{-42}$ & 708.3 & $2.354 \cdot 10^{5}$ & 0.7099 & 0.9708 & 0.7516 & $1.713 \cdot 10^{48}$ & $1.535 \cdot 10^{-7}$ & $1.972 \cdot 10^{51}$ & $1.767\cdot 10^{-4}$\\ 
        $3\cdot10^{-3}$ & $10^{15}$ & 100 & $10^{-41}$ & 568380 & $1.895 \cdot 10^{8}$ & 2.3761 & 1.0000 & 0.7516 & $5.416 \cdot 10^{48}$ & $4.854 \cdot 10^{-7}$ & $5.733 \cdot 10^{55}$ & 5.138 \\ 
        $3 \cdot 10^{-3}$ & $10^{15}$ & 100 & $10^{-40}$ & 51055 & $1.702 \cdot 10^{7}$ & 4.1838 & 1.0000 & 0.7516 & $1.713 \cdot 10^{49}$ & $1.535 \cdot 10^{-6}$ & $4.461 \cdot 10^{55}$ & 3.998\\ 
        $3\cdot10^{-3}$ & $10^{15}$ & 300 & $10^{-41}$ & 138607 & $4.620 \cdot 10^{7}$ & 1.3895 & 1.0000 & 0.7516 & $6.888 \cdot 10^{47}$ & $1.852 \cdot 10^{-7}$ & $3.741 \cdot 10^{54}$ & 1.006 \\ 
        $3\cdot10^{-3}$ & $10^{15}$ & 1000 & $10^{-41}$ & 113188 & $3.773 \cdot 10^{7}$ & 1.2924 & 1.0000 & 0.7516 & $9.498 \cdot 10^{46}$ & $8.513 \cdot 10^{-8}$ & $1.782 \cdot 10^{54}$ & 1.597 \\ 
        $3\cdot10^{-3}$ & $10^{15}$ & 3000 & $10^{-41}$ & 698 & $2.320 \cdot 10^{5}$ & 0.7076 & 0.9700 & 0.7506 & $1.482 \cdot 10^{46}$ & $3.985 \cdot 10^{-8}$ & $7.645 \cdot 10^{48}$ & $2.056 \cdot 10^{-5}$ \\ 
        $3\cdot10^{-3}$ & $10^{16}$ & 100 & $10^{-41}$ & 50818 & $1.694 \cdot 10^{7}$ & 4.1855 & 1.0000 & 0.7516 & $1.713 \cdot 10^{49}$ & $1.535 \cdot 10^{-6}$ & $4.404 \cdot 10^{55}$ & 3.947\\ 

        \hline
    \end{tabular}
    }
    \label{tab:models}
\end{table*}

\section{Results}\label{Sec:Results}

\subsection{Impact of WIMP mass}\label{Sec:mass}
In this section, we keep fixed the spin-dependent scattering cross section at $\sigma_{\rm SD} = 10^{-41}~\mathrm{cm^2}$ and the ambient dark matter density at $\rho_\chi = 10^{15}~\mathrm{GeV~cm^{-3}}$, varying only the WIMP mass $m_\chi$ with values $m_\chi \in \{30, 50, 100, 300, 1000, 3000\}~\mathrm{GeV}$. The accretion rate of baryonic matter is held constant at $3\times10^{-3} \, M_\odot \, {\rm yr}^{-1}$.


\begin{figure*}[h]
    \centering
    \includegraphics[width=0.49\textwidth]{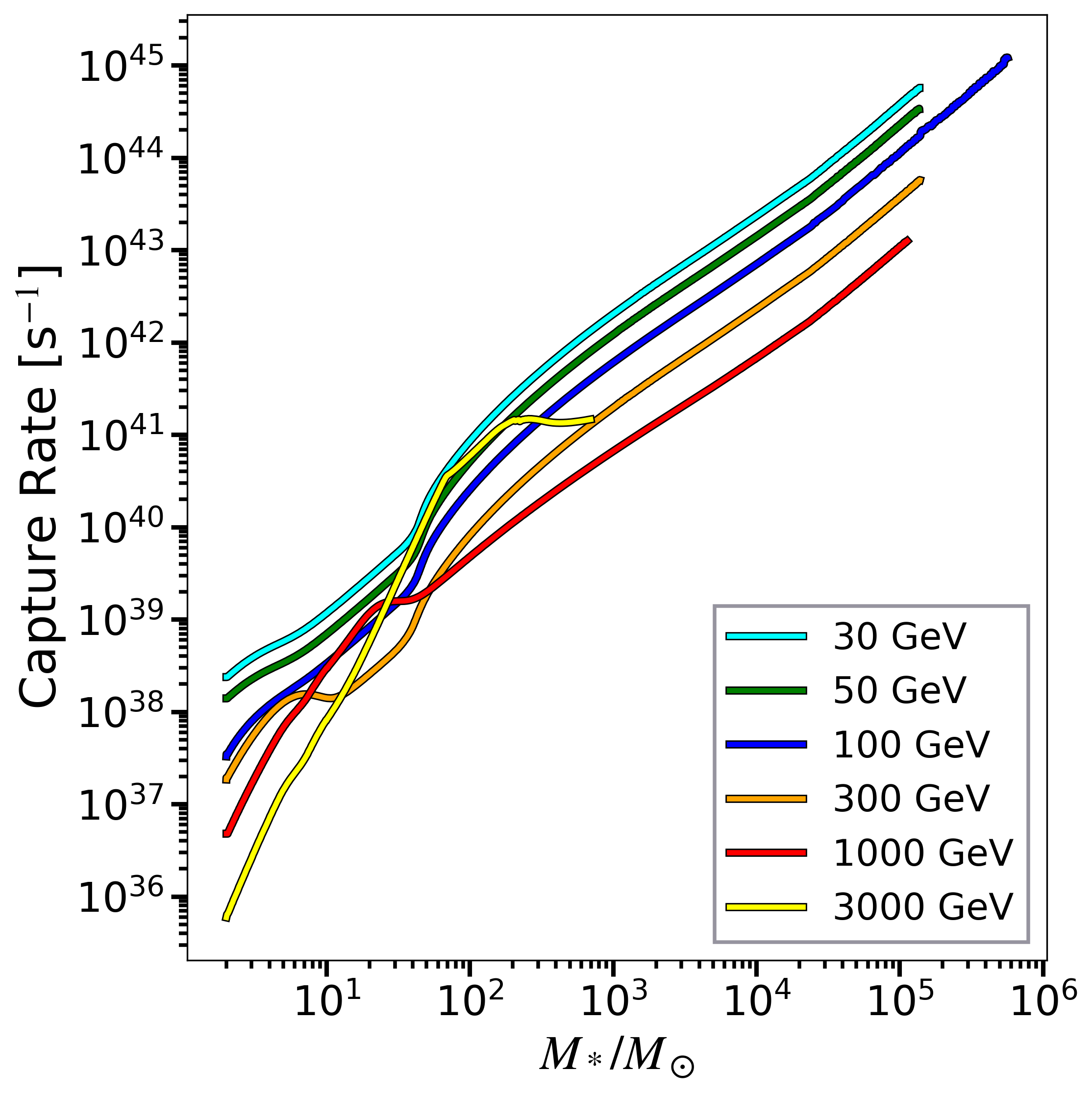}
    \includegraphics[width=0.49\textwidth]{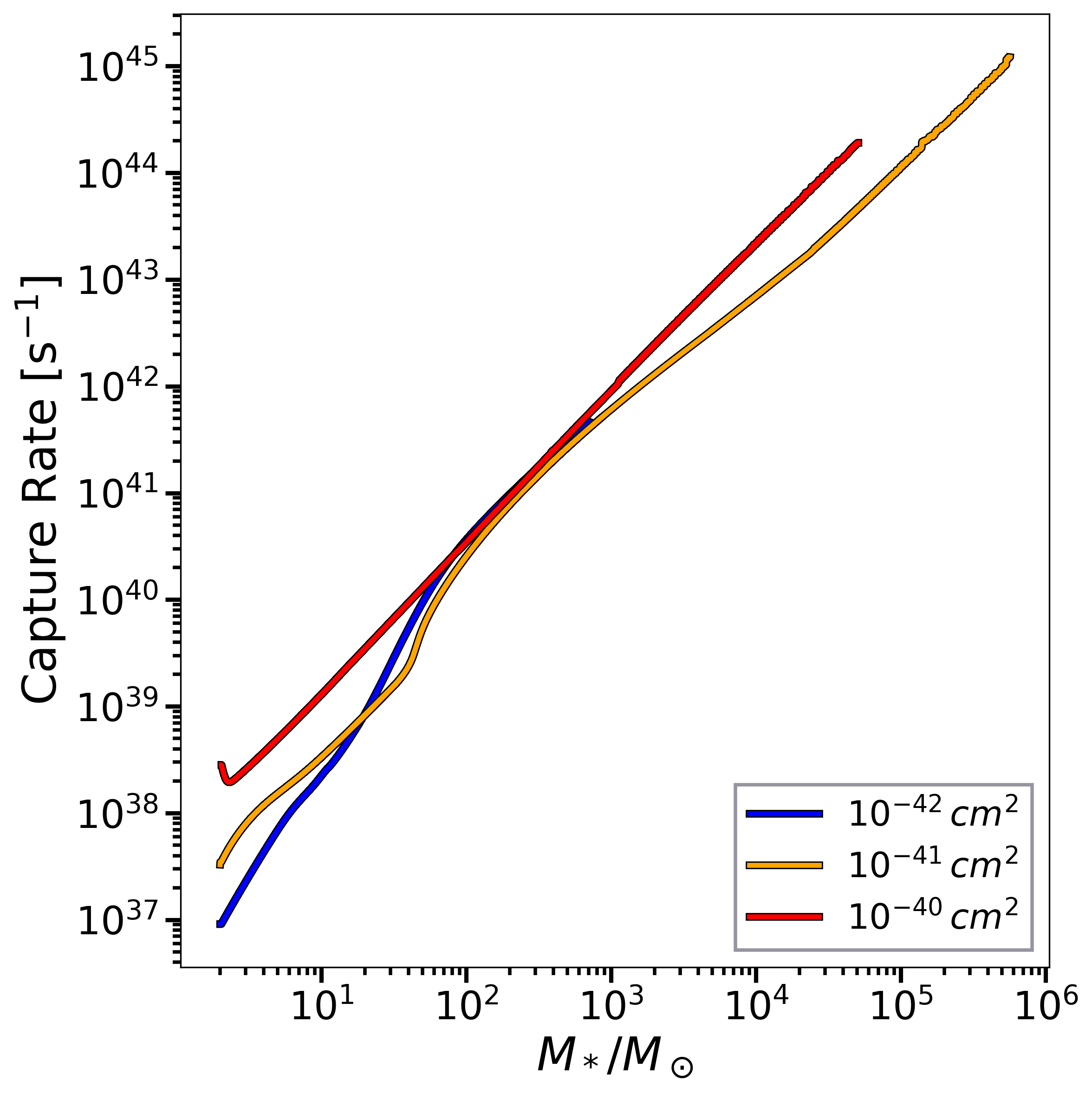}
    \caption{Total WIMPs capture rate $C_c$ as a function of stellar mass for accreting Population III protostar models. \textit{Left:} Six models at fixed ambient dark matter density $\rho_\chi = 10^{15} \, {\rm GeV \, cm}^{-3}$, accretion rate $\dot{M} = 3 \times 10^{-3} \, M_\odot \, {\rm yr}^{-1}$, and scattering cross section $\sigma_{\rm SD} = 10^{-41} \, {\rm cm}^2$, for WIMP masses $m_\chi =$ 30, 50, 100, 300, 1000, and 3000~GeV. \textit{Right:} Three models with fixed WIMP mass $m_\chi = 100 \, GeV$ and ambient dark matter density $\rho_\chi = 10^{15} \, {\rm GeV \, cm}^{-3}$, varying spin-dependent scattering cross section $\sigma_{\rm SD} = 10^{-42}$, $10^{-41}$, and $10^{-40} \, {\rm cm}^2$ for the same accretion rate.}
    \label{fig:CaptureRates}
\end{figure*}

Figure~\ref{fig:CaptureRates} (left panel) shows the evolution of the total WIMP capture rate, $C_{\rm c}(t)$. At the beginning of the protostellar phase, the capture rates are approximately $2\times10^{38}~\mathrm{s^{-1}}$ for $m_\chi = 30~\mathrm{GeV}$, $1.5\times10^{38}~\mathrm{s^{-1}}$ for 50 GeV, $3\times10^{37}~\mathrm{s^{-1}}$ for 100 GeV, $2\times10^{37}~\mathrm{s^{-1}}$ for 300 GeV,  $5\times10^{36}~\mathrm{s^{-1}}$ for 1000 GeV, and $5\times10^{35}~\mathrm{s^{-1}}$ for 3000~GeV. The first five models exhibit a monotonic increase in $C_{\rm c}(t)$ throughout the protostellar lifetime (up to $10^8~\mathrm{yr}$), reflecting the continued growth of both the stellar escape speed and the enclosed baryonic mass. The 3000 GeV model increases its capture rate at a steeper rate than the other models, until it reaches approximately $200 \, M_\odot$. It then stabilizes, finding a plateau around $10^{41}~\mathrm{s^{-1}}$ until the end of its evolution at $698\, M_\odot$.

The differences in capture rates across WIMP masses can be largely attributed to the dependence of $C_{\rm c}$ on the inverse of the WIMP mass, as implied by Eq.~\ref{eq:dcdm}. As a result, at any given time, lighter WIMPs are captured more efficiently. By the end of the protostellar phase, capture rates reach values on the order of $\sim10^{45}~\mathrm{s^{-1}}$ for the lightest WIMPs and decrease steadily with mass down to $\sim10^{43}~\mathrm{s^{-1}}$ for $m_\chi = 1000~\mathrm{GeV}$, spanning roughly two orders of magnitude. Thus, the WIMP capture rate increases steadily over the protostellar phase, with lighter WIMPs captured more efficiently, reflecting the inverse mass dependence.


\begin{figure*}[h]
    \centering
    \includegraphics[width=0.49\textwidth]{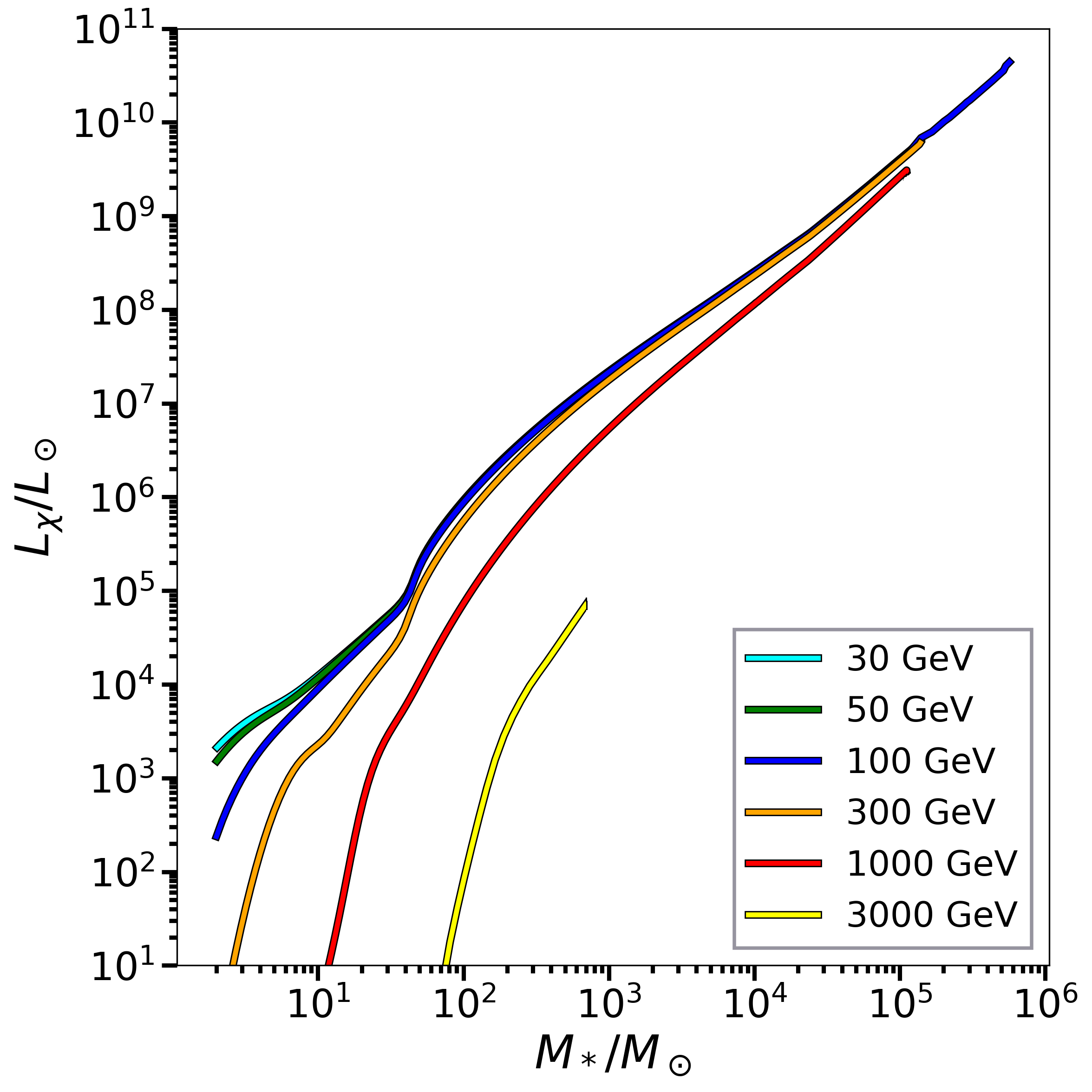}
    \includegraphics[width=0.49\textwidth]{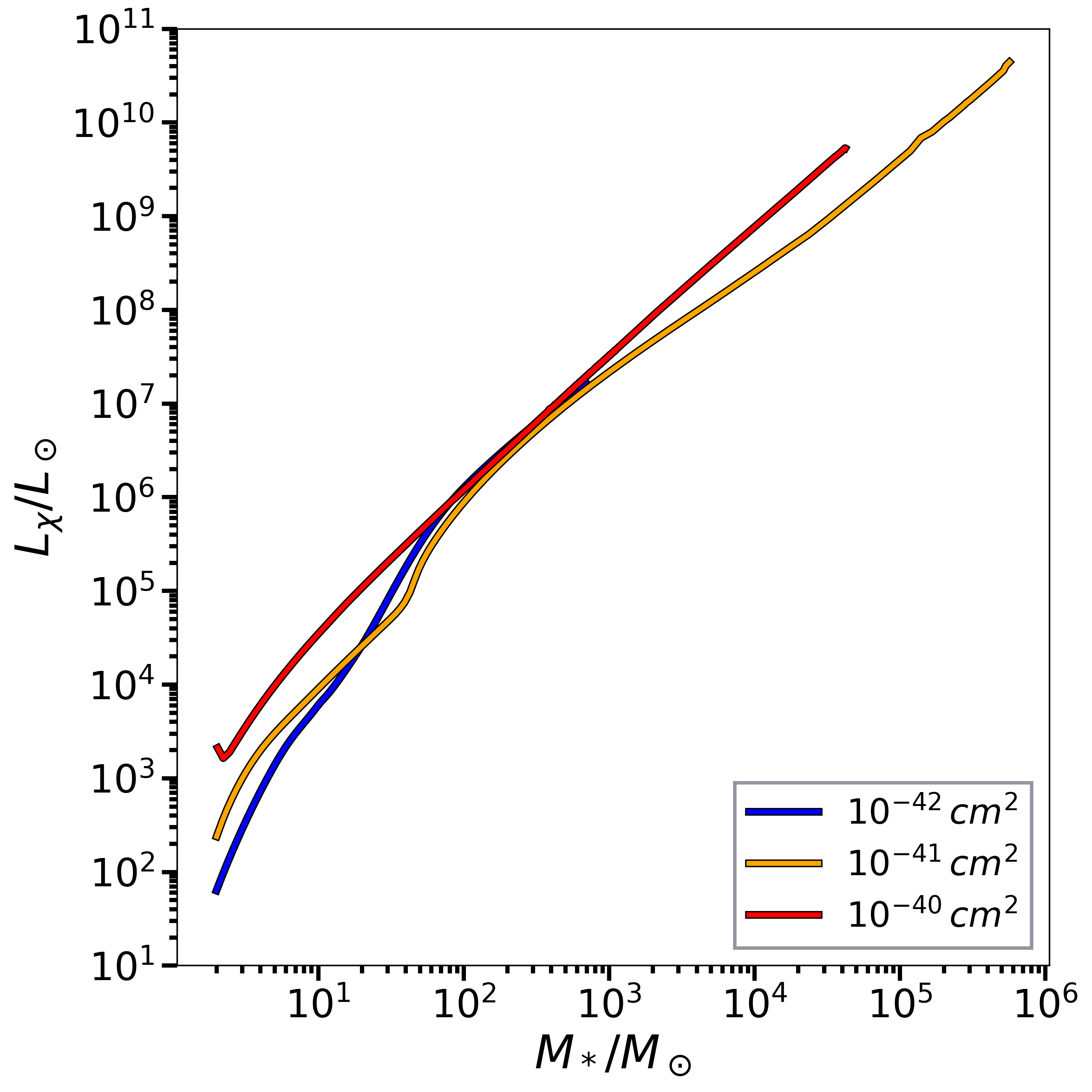}
    \caption{Annihilation luminosity $L_\chi = \frac{2}{3} \, m_\chi \, C_c$ as a function of stellar mass for accreting Population III protostar models. \textit{Left:} Six models at fixed ambient dark matter density $\rho_\chi = 10^{15} \, {\rm GeV \, cm}^{-3}$, accretion rate $\dot{M} = 3 \times 10^{-3} \, M_\odot \, {\rm yr}^{-1}$, and scattering cross section $\sigma_{\rm SD} = 10^{-41} \, {\rm cm}^2$, for WIMP masses $m_\chi =$ 30, 50, 100, 300, 1000 and 3000 GeV. \textit{Right:} Three models with fixed WIMP mass $m_\chi = 100 \, {\rm GeV}$ and ambient dark matter density $\rho_\chi = 10^{15} \, {\rm GeV \, cm}^{-3}$, varying spin-dependent scattering cross section $\sigma_{\rm SD} = 10^{-42}$, $10^{-41}$, and $10^{-40} \, {\rm cm}^2$ for the same accretion rate.}
    \label{fig:Lchi}
\end{figure*}

Figure~\ref{fig:Lchi} (left panel) presents the annihilation luminosity, $L_\chi(M_*) = \frac{2}{3} \, m_\chi \, C_{\rm c}$, as a function of the instantaneous stellar mass $M_*$, for the six different WIMP masses. The luminosities start at values of approximately $2\times10^3\,L_\odot$, $1.5\times10^3\,L_\odot$, and $2\times10^2\,L_\odot$ for the 30, 50, and 100 GeV models, respectively, while the 300, 1000, and 3000 GeV cases begin with negligible WIMP annihilation luminosity. As the star accretes mass from $2 \, M_\odot$ up to $\sim10^5 \, M_\odot$, all models exhibit a steep, monotonic rise in $L_\chi$. Ultimately they reach values in the range $10^9$–$10^{11}\,L_\odot$, depending on WIMP mass and the termination time of the computation, except for the 3000 GeV model.

At any given stellar mass, lighter WIMPs yield higher annihilation luminosities, reflecting the inverse dependence of the capture rate on $m_\chi$. This hierarchy persists throughout the evolution: the 30 and 50 GeV curves remain closely aligned from the beginning; the 100 GeV track joins them around $M_* \sim 10\,M_\odot$; the 300 GeV model converges near $M_* \sim 10^3\,M_\odot$; the 1000 GeV case consistently remains below the others, never fully catching up; and the 3000 GeV case remains even lower throughout its whole evolution, exhibiting a significant difference compared to all the other models.

The annihilation luminosity increases steeply as the star accretes mass, with lighter WIMPs producing higher luminosities at any given stellar mass. This mass hierarchy is maintained throughout the evolution, with the 1000 and 3000~GeV cases lagging behind the others.


\begin{figure*}[h]
    \centering
    \includegraphics[width=0.49\textwidth]{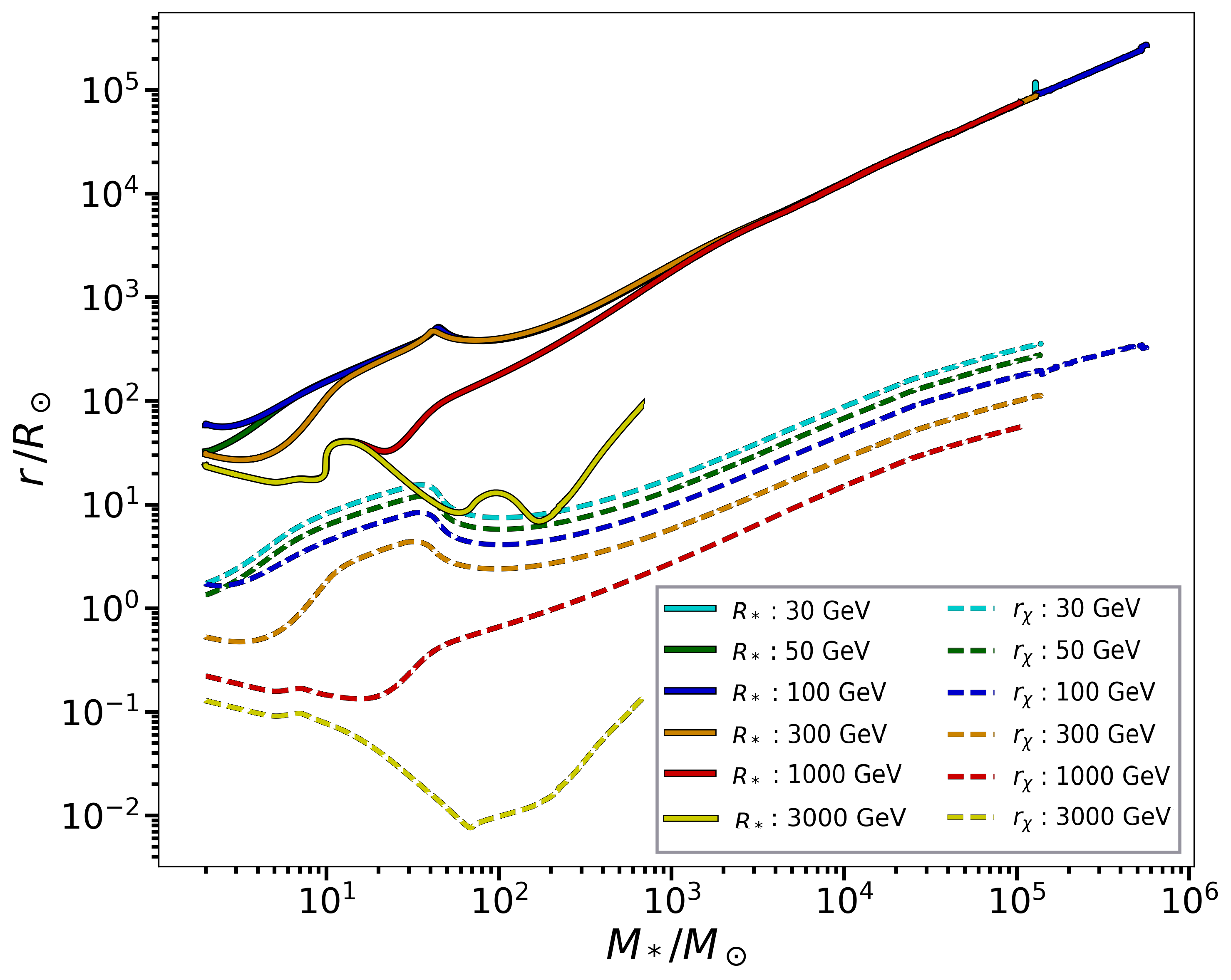}
    \includegraphics[width=0.49\textwidth]{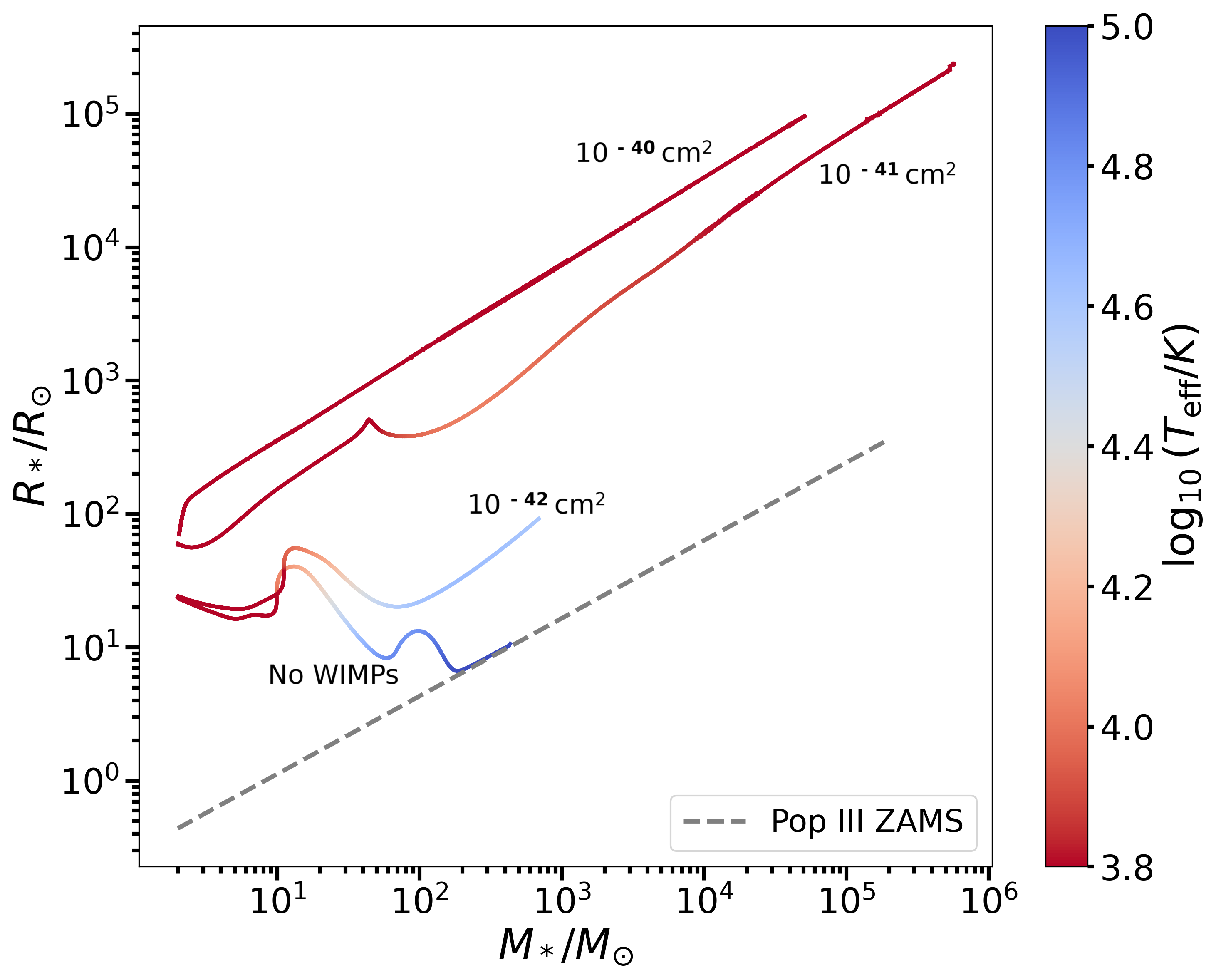}
    \caption{Radius versus stellar mass for accreting Population III protostar models. \textit{Left:} $R_*$ is the stellar radius and $r_\chi$ the WIMP characteristic length for six models at fixed ambient dark matter density $\rho_\chi = 10^{15} \, {\rm GeV \, cm}^{-3}$, accretion rate $\dot{M} = 3 \times 10^{-3} \, M_\odot \, {\rm yr}^{-1}$, and scattering cross section $\sigma_{\rm SD} = 10^{-41} \, {\rm cm}^2$, for WIMP masses $m_\chi =$ 30, 50, 100, 300, 1000 and 3000~GeV. \textit{Right:} Four models at a fixed WIMP mass $m_\chi = 100 \, GeV$ and ambient dark matter density $\rho_\chi = 10^{15} \, {\rm GeV \, cm}^{-3}$, for spin-dependent scattering cross sections $\sigma_{\rm SD} = 10^{-42}$, $10^{-41}$, $10^{-40} \, {\rm cm}^2$ and the No-WIMPs model.}
    \label{fig:Radii}
\end{figure*}

Figure~\ref{fig:Radii} (left panel) shows, as a function of instantaneous stellar mass $M_*$, both the stellar radius $R_*$ (solid lines) and the characteristic WIMP scale radius $r_\chi$ (dashed lines). All models begin at $M_*=2\,M_\odot$ with $20 \, R_\odot \leq R_*\leq 60 \, R_\odot$ and $0.1 \, R_\odot \leq r_\chi \leq 2 \, R_\odot$. As accretion proceeds to $M_*\sim10^3\,M_\odot$, $R_*$ rises to a few thousand $R_\odot$, exhibiting secondary inflationary loops in the low-mass WIMP cases (30 – 100 GeV). Beyond $M_*\sim10^3\,M_\odot$, all tracks (except for the 3000 GeV) converge onto a common power-law $R_*\propto M_*^{0.761}$ up to the computational limit of $M_*\sim10^5\,M_\odot$, with final radii in the order of $10^5 \, R_\odot$. The WIMP scale radius $r_\chi$ remains systematically smaller, at the range of less than $10\% \, R_*$, with lighter WIMPs yielding slightly larger $r_\chi$. The $r_\chi$ follow the same pattern of the evolution of the total radius of the star.

The 3000 GeV model experiences pulsations that affect its radius due to the contraction to the ZAMS phase. After the model reaches the ZAMS at around $200 \, M_\odot$, the radius grows steadily until the end of its evolution at $698 \, M_\odot$. Its WIMP radius is confined to a small region, about $0.1\%$ of $R_*$. Its premature termination at $698 \, M_\odot$ due to the radiative feedback condition, does not allow us to perform a further analysis with a power law, as we do for the other models.

The stellar radius $R_*$ grows significantly during accretion, eventually following a common power-law scaling across all WIMP masses, while the WIMP scale radius $r_\chi$ remains much smaller throughout. The WIMP scale radius remains consistently smaller than the stellar radius, with lighter WIMPs yielding slightly larger $r_\chi$ values.


\begin{figure*}[h]
    \centering
    \includegraphics[width=0.49\textwidth]{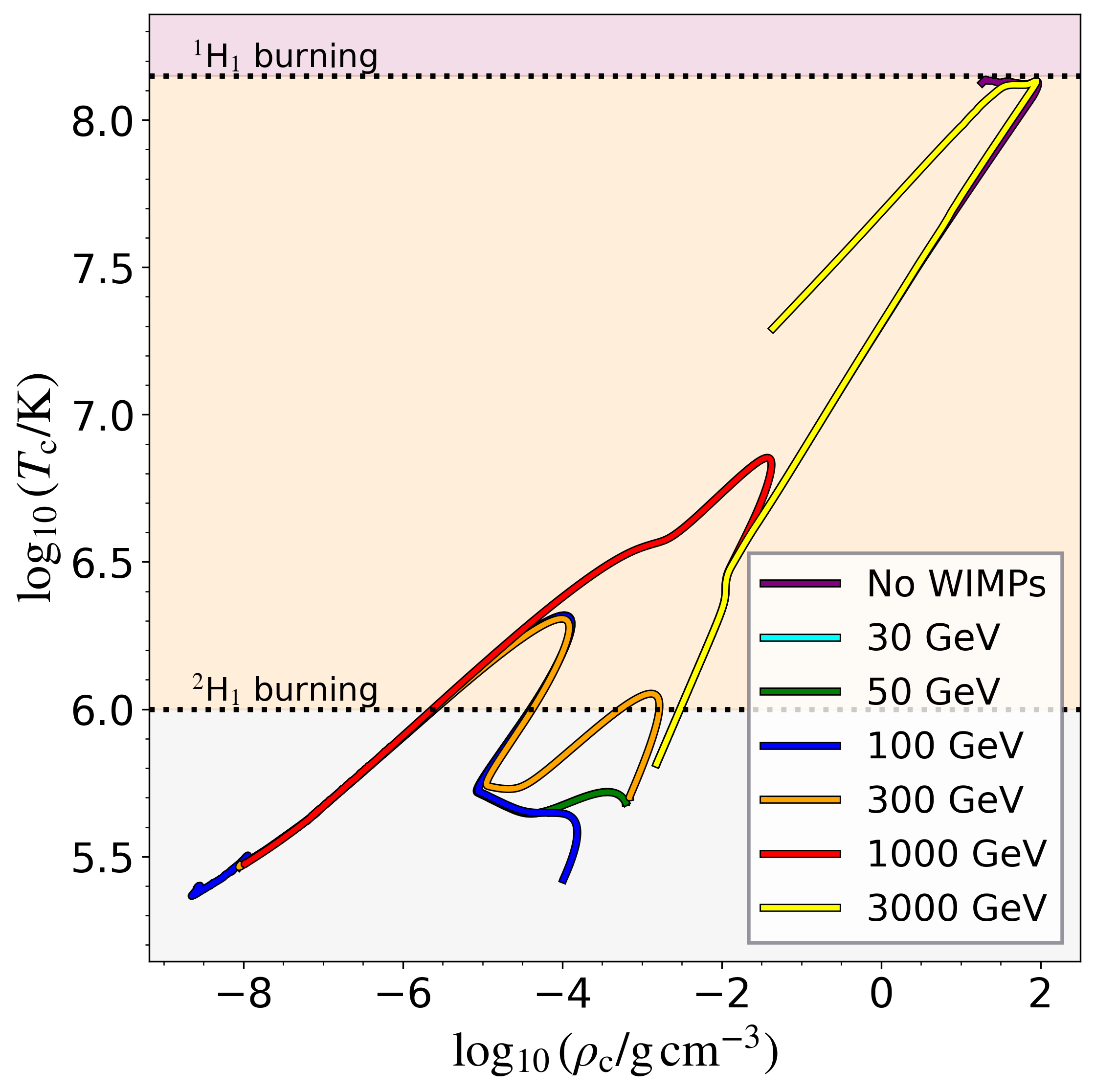}
    \includegraphics[width=0.49\textwidth]{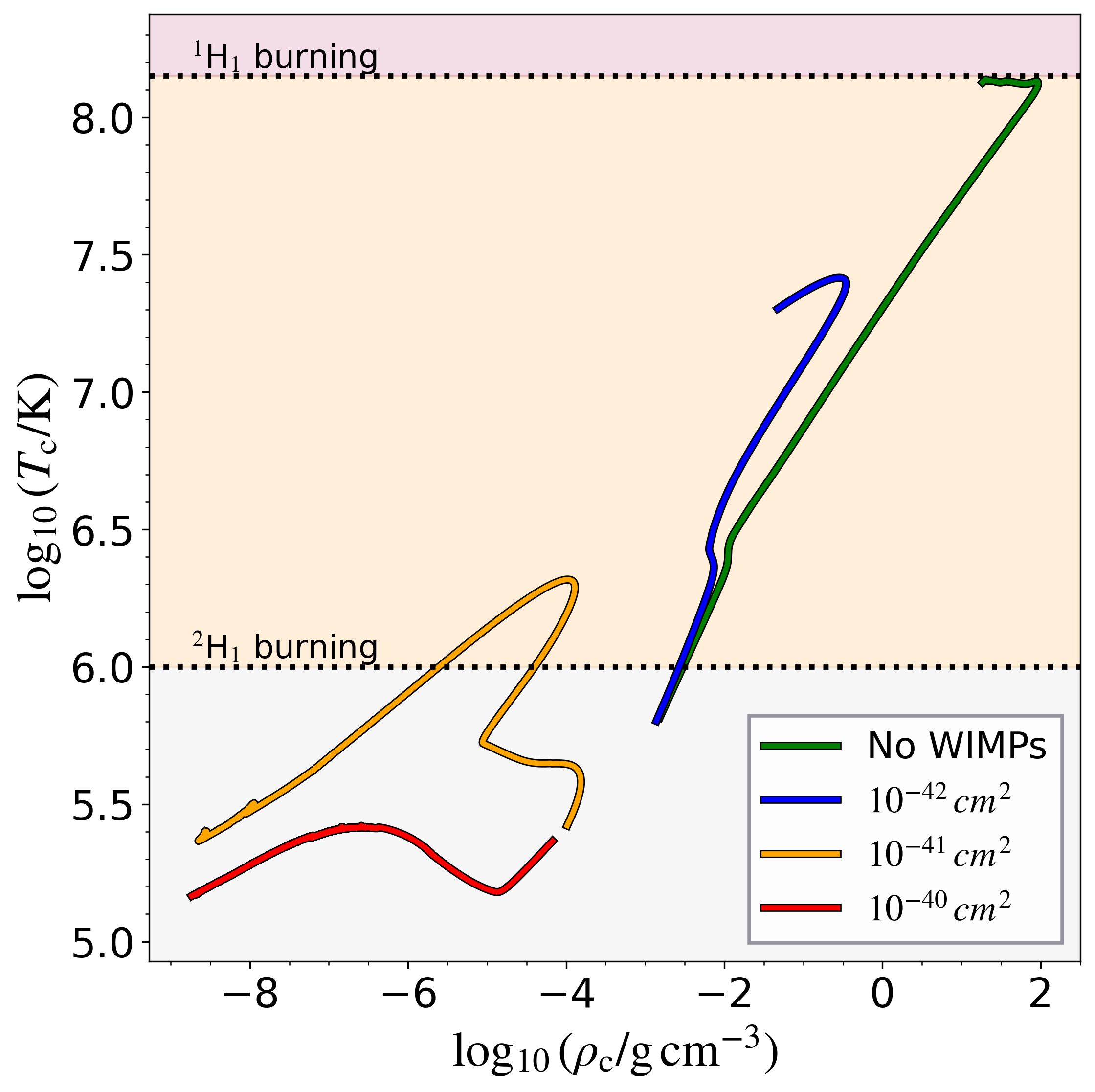}
    \caption{Evolution of central temperature versus central density for accreting Population III protostar models. \textit{Left:} Seven models at a fixed ambient WIMP density of $10^{15} \, {\rm GeV \, cm}^{-3}$ and accretion rate $\dot{M} = 3\times10^{-3} \, M_\odot \, {\rm yr}^{-1}$, for WIMP masses 30 GeV, 50 GeV, 100 GeV, 300, 1000, 3000~GeV, and the No-WIMPs case (standard Pop III). The colour-shaded regions denote regimes of no nuclear burning ($T_c < 10^6$ K), deuterium burning ($10^6 \lesssim T_c/{\rm K} < 10^{8.15}$), and hydrogen ignition ($T_c \gtrsim 10^{8.15}$ K), respectively. \textit{Right:} Four models at a fixed WIMP mass $m_\chi = 100$~GeV and ambient dark matter density $\rho_\chi = 10^{15} \, {\rm GeV \, cm}^{-3}$, for spin-dependent scattering cross sections $\sigma_{\rm SD} = 10^{-42}$, $10^{-41}$, $10^{-40} \, {\rm cm}^2$ and the No-WIMPs case. The colour-shaded regions represent the same nuclear burning temperatures.}
    \label{fig:Tcrhoc}
\end{figure*}


\begin{figure*}[h]
    \centering
    \includegraphics[width=0.49\textwidth]{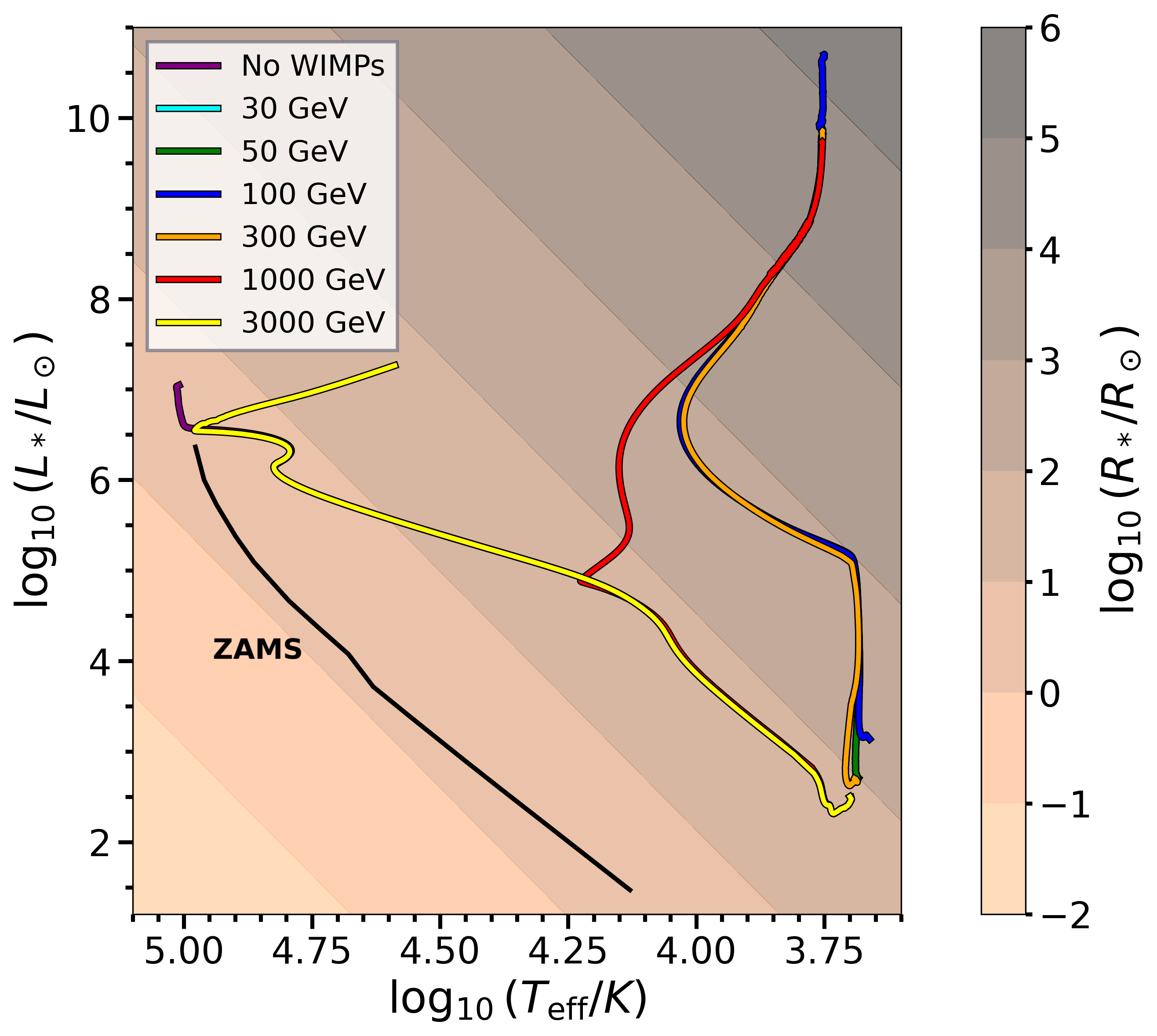}
    \includegraphics[width=0.49\textwidth]{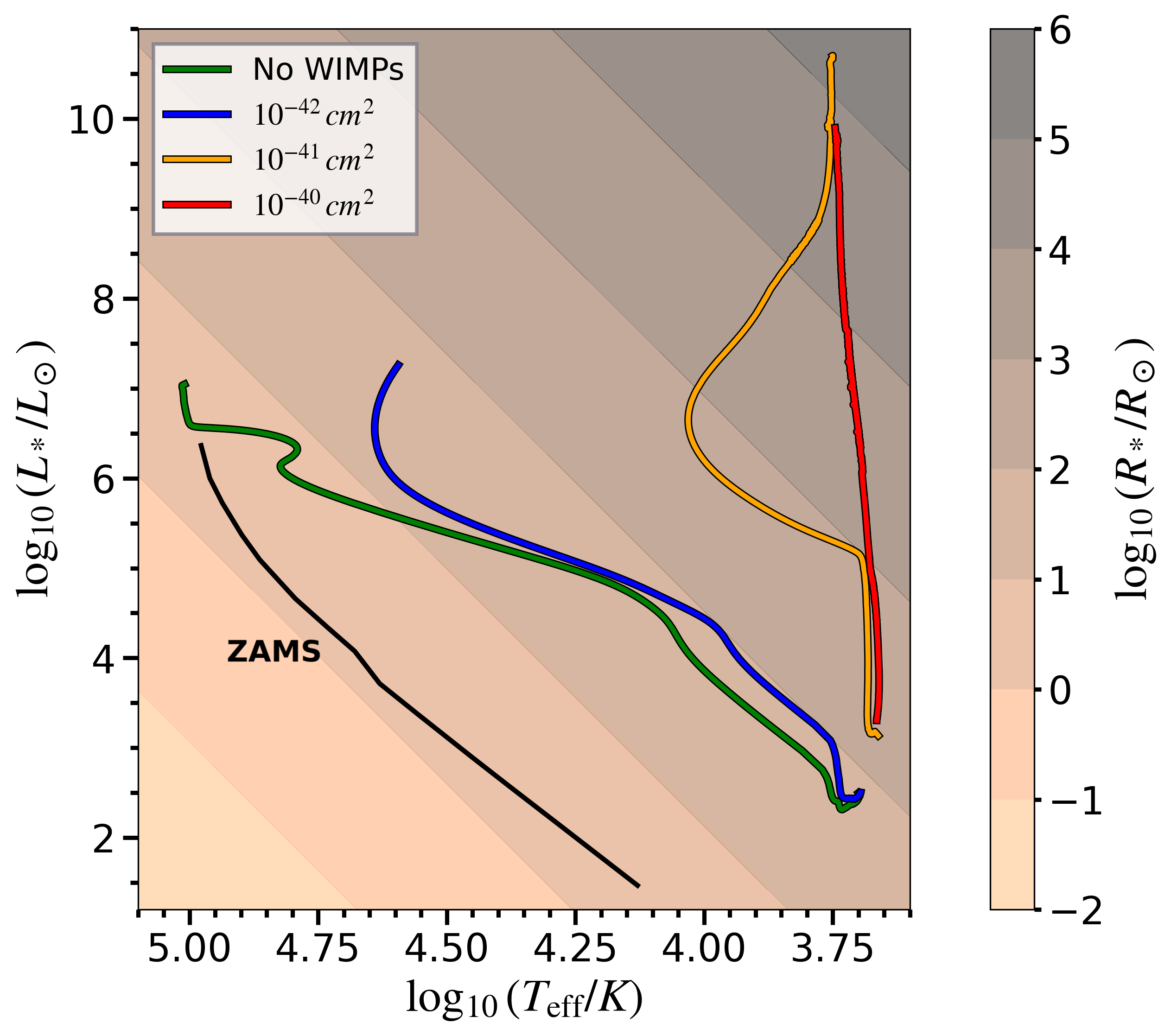}
    \caption{Hertzsprung–Russell diagrams for accreting Population III protostar models. \textit{Left:} Seven models at a fixed ambient WIMP density of $10^{15} \, {\rm GeV \, cm}^{-3}$ and accretion rate $\dot{M} = 3\times10^{-3} \, M_\odot \, {\rm yr}^{-1}$, for WIMP masses $m_\chi = $ 30 GeV, 50 GeV, 100 GeV, 300 GeV, 1000 and 3000 GeV and the No-WIMPs model (standard Pop III). The black line marks the Pop III Zero-Age Main Sequence (ZAMS). \textit{Right:} Four models at a fixed WIMP mass $m_\chi = 100$ GeV and ambient dark matter density $\rho_\chi = 10^{15} \, {\rm GeV \, cm}^{-3}$, for spin-dependent scattering cross sections $\sigma_{\rm SD} = 10^{-42}$, $10^{-41}$, and $10^{-40} \, {\rm cm}^2$ and the No-WIMPs case.}
    \label{fig:HRDs}
\end{figure*}

The No-WIMPs model (standard Pop III) contracts efficiently toward the zero-age main sequence (ZAMS), reaching $\log_{10}(T_{\mathrm{eff}}/{\rm K}) \simeq 4.8$ and $\log_{10}(L/L_\odot) \simeq 6$ before stalling due to hydrogen ignition (Figure ~\ref{fig:Tcrhoc} and ~\ref{fig:HRDs} left panels). In contrast, models with nonzero WIMP masses deviate from this path at progressively earlier stages, evolving toward cooler effective temperatures and larger radii as annihilation heating becomes significant. Lower-mass WIMP models (30 and 50 GeV) loop toward the Hayashi line near $\log_{10}(T_{\mathrm{eff}}/{\rm K}) \simeq 3.8$, performing redward–blueward excursions around $\log_{10}(L/L_\odot) \simeq 4$, consistent with intermittent deuterium burning. The higher WIMP mass models (1000 and 3000 GeV) exhibit higher maximum central temperatures ($10^{6.85}$ K and $10^{8.15}$ K), high enough to initiate core deuterium burning and core hydrogen burning, respectively.

Figure~\ref{fig:HRDs} (left panel) shows the evolutionary tracks of accreting protostars in the Hertzsprung–Russell diagram (HRD) for all the five different WIMP masses and the No-WIMPs case. All models begin their evolution at $M_* = 2\,M_\odot$, near $\log_{10}(T_{\mathrm{eff}}/\mathrm{K}) \simeq 3.7$ and $\log_{10}(L/L_\odot) \simeq 2.8$, and initially follow a vertical rise along the Hayashi track as accretion proceeds.

The 1000 GeV model initially follows a trajectory nearly identical to the standard Pop III, owing to its relatively low capture and annihilation rates. However, it eventually joins the other WIMP-powered-stars tracks, settling near the Hayashi line. This delayed deviation is associated with a phase of sustained deuterium burning, supported by the model’s higher central temperatures and densities. This allow deuterium fusion to persist for several thousand years, longer than in the lower-mass WIMP cases. By the end of their evolution the 300 and 1000 GeV tracks evolve nearly vertically at $\log_{10}(T_{\mathrm{eff}}/{\rm K}) \lesssim 4.0$, with luminosities climbing to $\log_{10}(L/L_\odot) \gtrsim 9$. 

An exception occurs again for the 3000 GeV model, which follows an HRD track identical to that of the standard Pop III star without WIMPs. It contracts toward the ZAMS, nearly touching it, and then deviates, inflating with a redward excursion. Unlike the other models, it does not reach the Hayashi line due to the termination of its accretion, but it reaches a final luminosity of $\log_{10}(L/L_\odot) = 7.3$.

In summary, the No-WIMPs model contracts efficiently to the ZAMS, while WIMP-powered models deviate early, evolving toward cooler temperatures and larger radii due to annihilation heating. Lower-mass WIMPs show redward excursions near the Hayashi line. Higher-mass WIMPs (1000 and 3000 GeV models) maintain higher central temperatures and initially follow the standard Pop III evolution. The 1000 GeV model later evolves with nearly vertical tracks in the HR diagram and reaches the Hayashi track, while the 3000~GeV model never reaches it.


\begin{figure}[h]
    \includegraphics[width=\columnwidth]{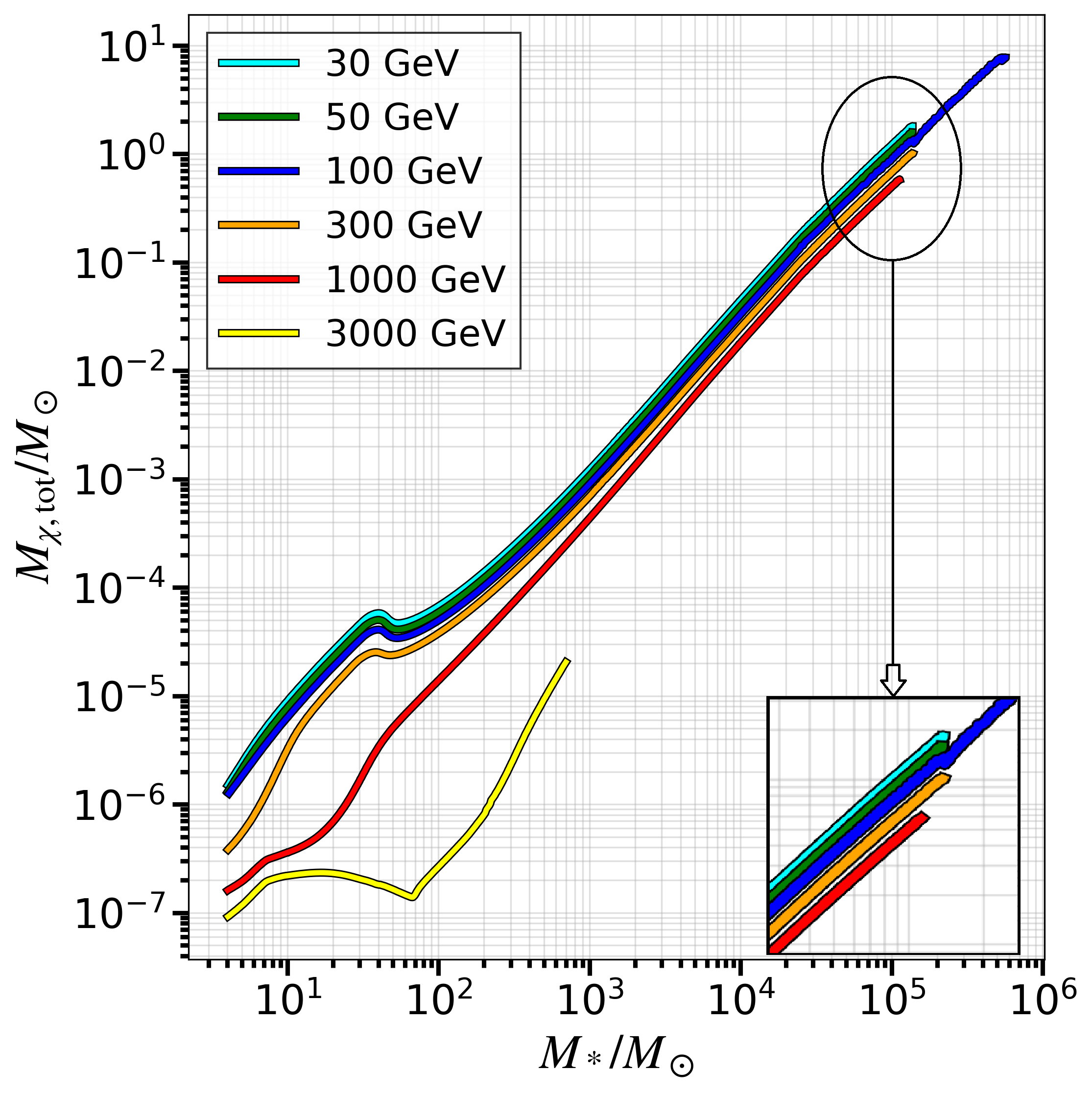}
    \caption{Total mass of WIMPs in the stars $M_{\chi,{\rm tot}}$ as a function of stellar mass $M_*$ for six accreting Population III protostar models at fixed ambient dark matter density $\rho_\chi = 10^{15} \, {\rm GeV \, cm}^{-3}$, accretion rate $\dot{M} = 3 \times 10^{-3} \, M_\odot \, {\rm yr}^{-1}$, and scattering cross section $\sigma_{\rm SD} = 10^{-41} \, {\rm cm}^2$, for WIMP masses $m_\chi =$ 30, 50, 100, 300, 1000 and 3000 GeV.}
    \label{fig:Mchitotal}
\end{figure}

Figure~\ref{fig:Mchitotal} plots the total mass of WIMPs in the star, $M_{\chi,{\rm tot}}(M_*) = N_\chi(M_*) \, m_\chi$, versus the stellar mass. The initial and final total WIMP masses of all the models are listed in Table~\ref{tab:models}. Across all WIMP masses, $M_{\chi,{\rm tot}}$ grows nearly as a power law, starting from approximately $500 \, M_\odot$. All the models are arranged in order, from lighter to heavier WIMPs, as dictated by Eq.~\ref{eq:Mchi}. Lighter WIMPs accumulate greater total WIMP masses compared to heavier WIMPs, reflecting the $1/m_\chi$ scaling of the capture rate. This ordering, with lower total DM masses for greater individual WIMP masses, also applies to the 3000 GeV model. However, a power law cannot be established for the 3000 GeV case, as it never reaches higher stellar masses.

At the evolutionary stage of $10^5 \, M_\odot$, all the models have approximately total WIMPs mass in the order of $1 \, M_\odot$. As aforementioned, lighter WIMPs exhibit greater total DM masses in the star. At $10^5 \, M_\odot$ the 30 GeV model has $\sim 1.32 \, M_\odot$, the 50 GeV $\sim 1.13 \, M_\odot$, the 100 GeV $\sim 1 \, M_\odot$, the 300 GeV $\sim 0.71 \, M_\odot$, and the 1000 GeV $\sim 0.55 \, M_\odot$.

\subsection{Impact of WIMP scattering cross-section}\label{Sec:sigmasd}

In this section, we keep fixed the WIMP mass at $m_\chi = 100$ GeV and ambient dark matter density at $\rho_\chi = 10^{15}$ GeV cm$^{-3}$, varying only the spin-dependent scattering cross section with values $\sigma_{\rm SD} \in \{10^{-42},\,10^{-41},\,10^{-40}\}~\mathrm{cm^2}$. The accretion rate of the baryonic matter is again $3\times10^{-3} \, M_\odot \, {\rm yr}^{-1}$.


Figure~\ref{fig:CaptureRates} (right panel) displays the total WIMP capture rate $C_{\rm c}$ as a function of instantaneous stellar mass $M_*$ for all the different $\sigma_{\rm SD}$ models. In all three cases, the capture rate increases approximately as $C_{\rm c} \propto M_*^2 / R_*$ (as was derived in Eq.~\ref{eq:capture}), rising from $C_{\rm c} \sim 10^{37}$–$10^{38}$ s$^{-1}$ at $M_* \simeq 2\,M_\odot$ to $C_{\rm c} \sim 10^{44}$ s$^{-1}$ by $M_* \simeq 10^5\,M_\odot$. The ordering by cross section is preserved across the full mass range: the largest capture rates correspond to the highest $\sigma_{\rm SD}$, with the $10^{-40}$ cm$^2$ model exceeding the $10^{-41}$ cm$^2$ and $10^{-42}$ cm$^2$ cases at all masses.

Power-law fits to the $C_{\rm c}(M_*)$ relations reveal mass scalings of $C_{\rm c} \propto M_*^{1.282}$ for $\sigma_{\rm SD} = 10^{-42} \,\mathrm{cm}^2$, $C_{\rm c} \propto M_*^{1.380}$ for $\sigma_{\rm SD} = 10^{-41} \,\mathrm{cm}^2$, and $C_{\rm c} \propto M_*^{1.415}$ for $\sigma_{\rm SD} = 10^{-40} \,\mathrm{cm}^2$. The total WIMP capture rate $C_{\rm c}$ increases with stellar mass for all spin-dependent cross-section models, roughly following $C_{\rm c} \propto M_*^2 / R_*$. Higher cross-section values yield consistently larger capture rates, and power-law fits reveal a steepening mass dependence with increasing $\sigma_{\rm SD}$.


Figure~\ref{fig:Lchi} (right) shows the corresponding WIMP annihilation luminosity, $L_\chi = \tfrac{2}{3} \, m_\chi \, C_{\rm c}$, plotted against $M_*$. All models exhibit a rapid increase in $L_\chi$, from $L_\chi \sim 10^2$–$10^3\,L_\odot$ at $M_* \simeq 2\,M_\odot$ to $L_\chi \sim 10^{9}$–$10^{10}\,L_\odot$ at $M_* \simeq 10^5\,M_\odot$. As with the capture rates, $L_\chi$ scales monotonically with $\sigma_{\rm SD}$, with the $10^{-40}$ cm$^2$ model producing the highest luminosity at all stages. The logarithmic separation between adjacent $\sigma_{\rm SD}$ tracks remains roughly constant across the entire mass range.

The WIMP annihilation luminosity $L_\chi$ rises rapidly with stellar mass, reaching $\sim 10^{10}\,L_\odot$ at $M_* \simeq 10^5\,M_\odot$. As with capture rates, $L_\chi$ scales consistently with the spin-dependent cross-section $\sigma_{\rm SD}$, with higher cross-sections producing higher luminosities and maintaining a roughly constant logarithmic separation across the mass range.


According to the Figure ~\ref{fig:Radii} (right panel) the No-WIMPs and $\sigma_{\rm SD} = 10^{-42}$ cm$^2$ models begin with initial radii of approximately $25\,R_\odot$. Both experience contraction phases: the No-WIMPs model contracts to the ZAMS after undergoing several radial pulsations, ultimately reaching a radius of about $10\,R_\odot$. The $10^{-42}$ cm$^2$ model also exhibits smaller pulsations but does not reach the ZAMS, stabilizing at a larger final radius near $90\,R_\odot$. In contrast, the $10^{-41}$ and $10^{-40}$ cm$^2$ models start with significantly larger radii, around $60\,R_\odot$ and $70\,R_\odot$, respectively. Both undergo overall radius expansion, with the $10^{-41}$ cm$^2$ model experiencing a brief contraction lasting roughly $10^4$ years. Ultimately, these two models reach very large radii on the order of $10^{5}\,R_\odot$. Power-law fits to the stellar radius–mass relation yield scalings of $R_* \propto M_*^{0.714}$ for $\sigma_{\rm SD} = 10^{-42} \,\mathrm{cm}^2$, $R_* \propto M_*^{0.770}$ for $\sigma_{\rm SD} = 10^{-41} \,\mathrm{cm}^2$, and $R_* \propto M_*^{0.645}$ for $\sigma_{\rm SD} = 10^{-40} \,\mathrm{cm}^2$.


Figure ~\ref{fig:Tcrhoc} (right panel) shows the evolution of the central temperature and density and combined with the HRD tracks of Figure ~\ref{fig:HRDs} (right panel) can provide valuable information. The stellar models begin their evolution along the HRD at slightly different positions. The $\sigma_{\rm SD} = 10^{-42}$ cm$^2$ and the No-WIMPs cases start near $\log_{10}(T_{\rm eff}/{\rm K}) \simeq 3.7$ and $\log_{10}(L/L_\odot) \simeq 2.5$, while the $10^{-41}$ and $10^{-40}$ cm$^2$ cases begin somewhat cooler and more luminous, around $\log_{10}(T_{\rm eff}/{\rm K}) \simeq 3.65$ and $\log_{10}(L/L_\odot) \simeq 3$.

As these protostars grow in mass, their evolutionary paths diverge based on the value of $\sigma_{\rm SD}$. Lower cross section models evolve toward the hotter, blue side of the HRD, consistent with contraction and progression toward the ZAMS. In contrast, models with higher $\sigma_{\rm SD}$ remain on the cooler, red side, with extended excursions along the Hayashi track due to WIMP-induced inflation.

As discussed in the previous section, the No-WIMPs model contracts efficiently to the ZAMS, reaching central temperatures high enough to ignite hydrogen and central densities of $\log_{10}(\rho_{\rm c}/\mathrm{g \, cm}^{-3}) \sim 2$. The $10^{-42}$ cm$^2$ model follows a similar path but stalls before reaching the ZAMS, achieving a peak central temperature of $\log_{10}(T_{\rm c}/{\rm K}) \simeq 7.5$, sufficient only for sustained deuterium burning. The $10^{-41}$ cm$^2$ model undergoes a brief contraction and deuterium-burning phase, then remains inflated near the Hayashi limit. Finally, the $10^{-40}$ cm$^2$ case evolves entirely along the Hayashi track, remaining cool both internally and at the surface, and never reaching temperatures required for nuclear burning.

So, stellar evolution in the HR diagram depends strongly on the spin-dependent cross-section $\sigma_{\rm SD}$. Lower cross-section models ($\sigma_{\rm SD} = 10^{-42} \,\mathrm{cm}^2$) and the standard Pop III (No-WIMPs) cases evolve toward the ZAMS, reaching higher temperatures and densities suitable for hydrogen or deuterium burning. In contrast, higher $\sigma_{\rm SD}$ models remain cool and inflated along the Hayashi track due to dominant WIMP heating, never reaching conditions required for sustained nuclear fusion.


\begin{figure}[h]
    \includegraphics[width=\columnwidth]{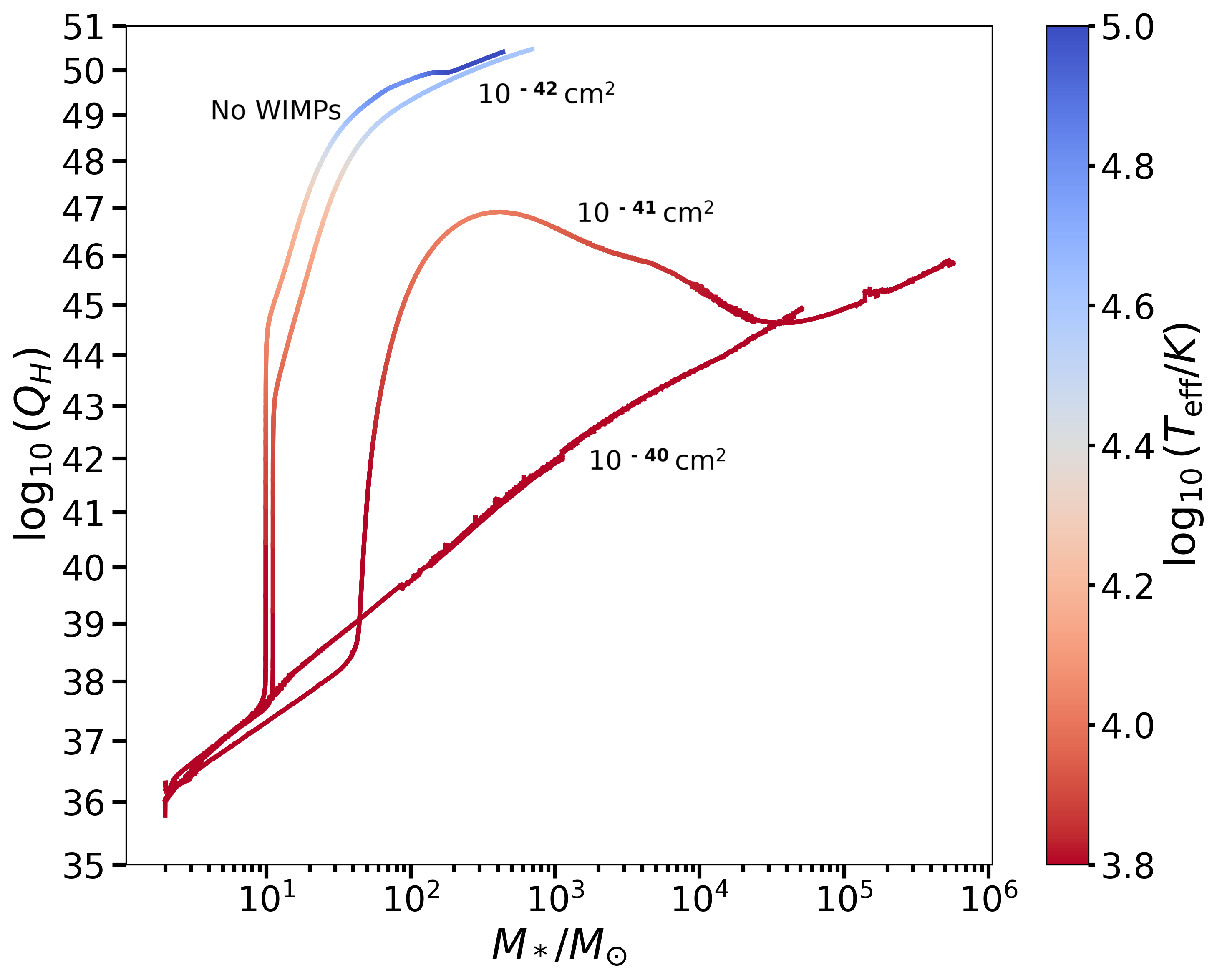}
    \caption{Hydrogen-ionizing photon production rate $Q_H$ as a function of stellar mass for four accreting Population III protostar models with fixed WIMP mass $m_\chi = 100 \, {\rm GeV}$ and ambient dark matter density $\rho_\chi = 10^{15} \, {\rm GeV \, cm}^{-3}$, for spin-dependent scattering cross sections $\sigma_{\rm SD} = 10^{-42}$, $10^{-41}$, and $10^{-40} \, {\rm cm}^2$ and the No-WIMPs case (standard Pop III).}
    \label{fig:QH_sigmasd}
\end{figure}

Regarding the hydrogen-ionizing photon production rate $Q_{\rm H}$, Figure ~\ref{fig:QH_sigmasd}, the contraction phases of all models correspond to elevated ionizing fluxes. The No-WIMPs and $10^{-42}$ cm$^2$ models produce the highest photon rates, reaching nearly $10^{51}$ s$^{-1}$. The $10^{-41}$ cm$^2$ model attains significantly lower peak values, up to about $10^{47}$ s$^{-1}$, while the $10^{-40}$ cm$^2$ model shows a steady increase in $Q_{\rm H}$ from its initial value near $10^{36}$ s$^{-1}$ to approximately $10^{45}$ s$^{-1}$ by the end of the simulation.

\subsection{Impact of DM density}\label{Sec:rhochi}


The impact of the WIMP density has already been shown in the first paper of this series \citep{Nandal2025}. There we show stellar radius $R_*/R_\odot$ as a function of instantaneous mass $M_*/M_\odot$ for six ambient dark matter densities $\rho_\chi = 10^{12},\,10^{13},\,10^{14},\,5\times10^{14},\,10^{15},\,10^{16}$ GeV cm$^{-3}$ for WIMPs of mass $m_\chi = 100$ GeV and $\sigma_{\rm SD} = 10^{-41}$ cm$^2$, with a baryonic accretion rate $\dot{M} = 3\times10^{-3} \, M_\odot \, {\rm yr}^{-1}$.

The models with $\rho_\chi=10^{12}$ and $10^{13}$ GeV cm$^{-3}$ exhibit no significant deviation from the evolution of a standard Pop III protostar. They contract steadily toward the ZAMS, reaching $R_*\sim10\,R_\odot$ by $M_*\sim100\,M_\odot$, and ultimately terminate at $M_*\approx443\,M_\odot$ and $445\,M_\odot$, respectively, when radiative feedback halts accretion. The $\rho_\chi=10^{14}$ GeV cm$^{-3}$ model undergoes continued contraction but does not reach the ZAMS; instead, radiative feedback intervenes at $M_*=702\,M_\odot$. For $\rho_\chi=10^{15}$ GeV cm$^{-3}$, the star inflates monotonically up to $M_*=43\,M_\odot$, undergoes a modest contraction phase until $M_*=91\,M_\odot$, and subsequently re-inflates as it approaches the Hayashi track. The highest density case, $\rho_\chi=10^{16}$ GeV cm$^{-3}$, exhibits no appreciable contraction phase; the star inflates continuously, reaching $R_*\sim10^5\,R_\odot$ by the end of the computation at $M_*\approx5\times10^4\,M_\odot$.


\section{Discussion}\label{Sec:Discussion}

\subsection{Comparative Analysis of WIMP Masses}
\label{Sec:mchi}

Our results on the parametric impact of the WIMP mass, $m_\chi$, on the evolution of Pop III.1 protostars extend and refine previous studies of dark matter-powered star formation and early stellar mass assembly. We find that lighter WIMPs (30–50 GeV) induce rapid deviations from canonical protostellar contraction, driving excursions back toward the Hayashi track at lower effective temperatures and intermittent deuterium-flashing loops (Figs.~\ref{fig:Tcrhoc},~\ref{fig:HRDs}). In contrast, very heavy WIMPs ($m_\chi\gtrsim300$ GeV) delay the onset of annihilation heating, leading to more prolonged contraction and a subsequent, steeper inflation only once central temperatures and densities suffice for modest capture rates (Figs.~\ref{fig:CaptureRates},~\ref{fig:Lchi}).

Our qualitative finding that annihilation heating halts contraction and inflates the stellar radius mirrors the seminal work of \cite{Spolyar2008} who hypothesized "dark stars" powered by DM annihilation. In their one-zone quasi-homogeneous approximation, they showed that capture and annihilation can dominate over nuclear burning at $M_*\approx10 \text{ - }100 \, M_\odot$ for WIMP densities $\gtrsim10^{13}$ cm$^{-3}$. This critical number density can occur at greater ambient mass densities $\rho_\chi$ in our models, simply because the scattering cross-sections used are much lower, by up to 2 orders of magnitude, according to the most recent experimental constraints \citet{LZ2024}.

Our GENEC models confirm that light WIMPs indeed produce luminosities $L_\chi\sim10^7 \, L_\odot$ at similar mass scales of approximately $1000 \, M_\odot$ and drive Hayashi‐line tracks \citep{Ilie2012}. The primary difference from these models lies in our treatment of the ambient dark matter density: we assume a constant $\rho_\chi$, whereas they incorporate adiabatic contraction \citep{Blumenthal1986}. However, we find that for $m_\chi\gtrsim300$ GeV, the delay in heating leads to a bifurcated evolutionary path. Consistent with our findings, \citet{Freese2010} and \citet{RindlerDaller2015} report a similar ordering of HRD tracks with respect to WIMP mass, wherein models with heavier WIMPs exhibit systematically bluer (i.e., higher effective temperature) tracks compared to those with lighter WIMPs. However, direct comparison is limited, as the specific trajectories differ due to variations in the adopted ambient dark matter density $\rho_\chi$ and $\sigma_{\rm SD}$.

Our detailed computation of capture rates (Fig.~\ref{fig:CaptureRates}) and annihilation luminosities (Fig.~\ref{fig:Lchi}) as functions of instantaneous stellar mass highlights the well-known inverse dependence on WIMP mass $C_{\rm c} \propto 1/m_\chi$ \citep{Gould1987}. The resulting hierarchy of $L_\chi$ at fixed $M_*$ quantitatively agrees with the equilibrium state between captured and annihilated WIMPs. During contraction to ZAMS, the rising escape velocity dramatically boosts WIMP capture. If dark matter density is high enough, this may reignite or prolong a DM-powered star phase, or at least significantly influence the early ZAMS structure and evolution. Therefore, we can state that stellar contraction phases are associated with more efficient DM capture.

In our simulations the total captured DM mass grows similarly for WIMP masses ($m_\chi$) between 30 GeV and 3 TeV. At any given time the cumulative captured mass converges to nearly the same value across this range. Equivalently, the ``mass capture rate'' ($m_\chi C_{c}$) is approximately independent of $m_\chi$ (Eq.~\ref{eq:capture2}). This behaviour is fully consistent with efficient single-scatter capture when capture is dominated by deep interior shells of the star (in our case, the most efficient capture is approximately at the inner radius of $r_\chi$). In the deep-potential region the fractional kinetic-energy loss required for capture is small, so a single scatter can bind WIMPs over a broad mass range and the per-particle capture efficiency becomes effectively mass-independent. As a result, $m_\chi C_{c}$ increases and all the models converge, while the ``number capture rate'' $C_{c}$ scales as $1/m_\chi$, as shown in Fig.~\ref{fig:CaptureRates} (given that we hold the halo DM density fixed). We emphasize that observing $C_{c} \propto 1/m_\chi$ therefore does not by itself imply an optically-thick, multi-scatter regime \citep{Bramante2017}. Identical scalings can arise in single-scatter models whenever capture is concentrated in regions with sufficiently large local escape velocity.

A key finding is the identification of a threshold WIMP mass, $m_\chi\sim300$ GeV, above which dark-matter‐powered phases remain subdominant until after strong deuterium burning ($M_*\sim 10 \text{ - }100 \, M_\odot$). The 1000 GeV model’s later‐onset inflation further extends these results into the TeV regime, suggesting that only very heavy WIMPs allow protostars to approach ZAMS-like contraction before dark matter heating becomes important. Our models predict that to form ``heavy'' seeds, i.e., $\sim10^5 \, M_\odot$, for black hole formation, lower WIMP masses (below a few TeV) are required. At WIMP masses of 3 TeV, our fiducial simulations show that the radiative feedback condition is met, causing the stars to stop accreting at around $700 \, M_\odot$, resulting in ``light'' seeds.

The total dark matter mass in equilibrium, $M_{\chi, \, tot}$ , scales approximately as $m_\chi^{-1/4}$, as derived in Eq.~\ref{eq:Mchi}. This weak inverse dependence arises from the combination of thermal scale radius scaling and the capture-annihilation balance in the equilibrium regime. Physically, although heavier WIMPs contribute more mass per particle, their reduced number density and more compact thermal distribution lead to a smaller total mass bound within the star. For example, as can be observed in Fig.~\ref{fig:Mchitotal} in the zoomed-in panel, when $M_{\chi, \, tot} \approx 1\,M_\odot$ for $m_\chi = 100\,\mathrm{GeV}$, for the $m_\chi = 1000\,\mathrm{GeV}$ is $M_{\chi, \, tot} \approx 0.56\,M_\odot$, consistent with the scaling relation $M_{\chi,1000}/M_{\chi,100} \approx (1000/100)^{-1/4} \approx 0.56$. This trend is important when interpreting the contribution of DM heating in models spanning a wide WIMP mass range, as it implies that stars harboring very heavy WIMPs contain less total DM mass at equilibrium, despite the higher energy yield per annihilation event.

\subsection{Comparing the Spin-Dependent Scattering Cross-Sections}
\label{Sec:sigmasd}

In this study, our GENEC simulations reveal a clear bifurcation in protostellar evolution across the critical range $\sigma_{\rm SD} \sim 10^{-42}$–$10^{-41}$ cm$^2$, marking a transition between classical Pop III behavior and Pop III.1 star evolution. At low cross sections ($\sigma_{\rm SD} \lesssim 10^{-42}$ cm$^2$), capture and annihilation rates remain too low to significantly impact stellar structure and evolution. These stars contract efficiently and closely follow the evolution of standard Pop III protostars, eventually approaching the ZAMS. In fact, the $\sigma_{\rm SD} = 10^{-42}$cm$^2$ model with $\rho_\chi = 10^{15}$ GeV cm$^{-3}$ mimics the behavior of the $\sigma_{\rm SD} = 10^{-41}$cm$^2$ model at $\rho_\chi = 10^{14}$ GeV cm$^{-3}$. In both cases, radiative feedback conditions are met early, terminating accretion at approximately $708\,M_\odot$, which aligns with the formation of a 'light seed' black hole.

In contrast, models with $\sigma_{\rm SD} \geq 10^{-41}$ cm$^2$ exhibit strong annihilation heating from early stages, inflating the star along the Hayashi track and preventing contraction toward the ZAMS. These stars remain in extended, cool configurations for most of their evolution, allowing them to continue accreting up to final masses of several $\sim 10^5\,M_\odot$, producing potential 'heavy seeds' for black hole formation. The $\sigma_{\rm SD} = 10^{-40}$cm$^2$ case behaves similarly, with even stronger annihilation support and longer-lived inflation.

\begin{table}[t]
\centering
\caption{Comparison of theoretical and simulated capture rate exponents for different values of $\sigma_{\rm SD}$.}
\resizebox{\columnwidth}{!}{%
\begin{tabular}{|c|c|c|c|c|}
\hline
$\sigma_{\rm SD}\;(\mathrm{cm}^2)$ 
& $R_* \propto M_*^\beta$ 
& Theoretical $C \propto M_*^\alpha$ 
& Simulated $C \propto M_*^\alpha$ 
& \% Difference \\
\hline
$10^{-42}$ 
& 0.714 
& 1.286 
& 1.282 
& 0.31\% \\
\hline
$10^{-41}$ 
& 0.770 
& 1.230 
& 1.380 
& 12.20\% \\
\hline
$10^{-40}$ 
& 0.645 
& 1.355 
& 1.415 
& 4.43\% \\
\hline
\end{tabular}
}
\label{tab:sigmasd}
\end{table}

As described by Eq.~\ref{eq:capture}, the WIMP capture rate scales approximately as $C_{\rm c} \propto M_*^2 / R_*$ \citep[also in agreement with][]{Iocco2008}, and thus depends sensitively on the mass–radius relation. Our simulations outputs and theoretical predictions are displayed in the Table~\ref{tab:sigmasd}. The corresponding percentage differences between the theoretical and simulated power-law exponents are 0.31\% for $\sigma_{\rm SD} = 10^{-42}$ cm$^2$, 12.20\% for $\sigma_{\rm SD} = 10^{-41}$ cm$^2$, and 4.43\% for $\sigma_{\rm SD} = 10^{-40}$ cm$^2$. 

These results show excellent agreement at low and high cross sections, while the intermediate case exhibits a moderate deviation. The largest difference (12.20\%) at $\sigma_{\rm SD} = 10^{-41}$cm$^2$ reflects structural transitions during this critical phase, where strong annihilation heating causes rapid changes in internal stellar structure and breaks down the simple mass–radius power-law assumptions. Overall, the alignment supports the theoretical scaling derived from first principles, while highlighting the nonlinear feedback effects near the transition regime.

Together, these results identify $\sigma_{\rm SD} \sim 10^{-42}$–$10^{-41}$ cm$^2$ as a critical threshold: below it, protostars evolve toward canonical ZAMS tracks and light seed formation; above it, dark matter heating alters protostellar structure and feedback, enabling the formation of more massive objects. These findings bridge the gap between earlier capture-only analytic models \citep{Gould1987,Taoso2008} and full stellar evolution simulations \citep{RindlerDaller2015} that include adiabatic contraction, underscoring the crucial role of dark matter scattering physics in shaping the early Universe’s stellar populations.

\section{Conclusions}\label{Sec:Conclusions}

Based on the suite of GENEC stellar–evolution simulations presented in this work, which explore the three–dimensional grid in ambient WIMP density ($\rho_\chi=10^{12}-10^{16}\,\mathrm{GeV\,cm^{-3}}$), WIMP mass ($m_\chi=30-1000$ GeV) and spin–dependent scattering cross section ($\sigma_{\rm SD}=10^{-42}-10^{-40} \mathrm{ cm^2}$) for two representative baryonic accretion rates ($\dot M_*=10^{-3},\,3\times10^{-3}\,M_\odot\,\mathrm{yr^{-1}}$), we draw the following conclusions:

\begin{enumerate}

\item \textbf{WIMP mass threshold for ``heavy'' black hole seed formation.}
Our results demonstrate a clear dependence on WIMP mass for the formation of heavy black hole seeds. For WIMP masses below a few TeV (under the specific conditions set for ambient DM-density and scattering cross sections), DM heating enables the growth of massive protostars, allowing for the formation of black hole seeds on the order of $10^5 \, M_\odot$. In particular, models with WIMP masses of 3 TeV show a radiative feedback effect that truncates accretion at $\sim 700 \, M_\odot$, resulting in ``light'' seeds. These findings suggest that only WIMP masses within a specific range ($<3\:$TeV) are effective in producing the massive progenitors required for the formation of supermassive black holes in the early universe.

\item \textbf{Dependence on WIMP mass ($m_\chi$).}
Lighter WIMPs (30--50 GeV) are captured far more efficiently (capture rate $C_c\propto 1/m_\chi$ in the single–scatter formalism), producing higher annihilation luminosities $L_\chi$ at fixed stellar mass and thus producing earlier and stronger protostellar inflation. Heavier WIMPs ($m_\chi\gtrsim300$ GeV) delay the onset of annihilation–dominated phases: such models can contract closer to ZAMS–like states before WIMP heating becomes important. In equilibrium the total dark–matter mass bound in the star follows the weak scaling $M_{\chi,\rm tot}\propto m_\chi^{-1/4}$, so heavier WIMPs supply less total DM mass even though each annihilation releases more energy.

\item \textbf{Role of the spin–dependent cross section ($\sigma_{\rm SD}$) and a bifurcation in evolution.}
The grid shows a clear bifurcation around $\sigma_{\rm SD}\sim10^{-42}-10^{-41}\,\mathrm{cm^2}$. For $\sigma_{\rm SD}\lesssim10^{-42}\,\mathrm{cm^2}$ capture and annihilation are subdominant: protostars contract, approach the ZAMS, and produce large ionizing photon fluxes that terminate accretion at moderate stellar masses (`light seeds'). For $\sigma_{\rm SD}\gtrsim10^{-41}\,\mathrm{cm^2}$ annihilation heating dominates early, driving extended, cool Hayashi–track configurations with greatly reduced ionizing output and allowing growth to \(\sim10^{5}\,M_\odot\) or more (`heavy seeds').

\item \textbf{Scaling relations and structural response.}
Our simulations reproduce the approximate theoretical scaling $C_c\propto M_*^2/R_*$ (capture $\propto v_{\rm esc}^2$), and the fitted power–law exponents for $C_c(M_*)$ across the explored $\sigma_{\rm SD}$ values lie close to the theoretical expectation (simulated exponents $\simeq 1.282$, 1.38 and 1.415 for $\sigma_{\rm SD}=10^{-42},10^{-41},10^{-40}\,\mathrm{cm^2}$, respectively). The stellar radius–mass relation for inflated models follows an approximate power law, explaining why capture and luminosity rises steeply during growth and why inflation becomes self–sustaining in the high–capture regime.


\item \textbf{Robustness, assumptions and limitations.}
The qualitative conclusions above are robust across the explored grid, but they rest on a set of simplifying assumptions: (i) the ambient WIMP density is held constant (we do not include adiabatic contraction or halo depletion by the star), (ii) we adopt the single–scatter Gould capture formalism and neglect self–capture and evaporation (justified for $m_\chi\gtrsim30$ GeV and the cross–sections explored), (iii) the annihilation cross section $\langle\sigma_a v\rangle$ and $\sigma_{\rm SI}$ are fixed, and (iv) the models are one–dimensional and assume idealized constant mass–accretion rates. Each of these choices can quantitatively alter thresholds and timescales: for example, adiabatic contraction of the halo would generally modify the local $\rho_\chi$ and help sustain annihilation heating, while multi–D effects (rotation, fragmentation, disk shielding) could modify accretion geometries and feedback.

\end{enumerate}

Our results demonstrate a plausible, physically motivated pathway for forming heavy black hole seeds in the early Universe provided sufficiently dense DM environments are available. To make progress toward predictive, testable observational models we recommend the following extensions: (a) couple GENEC–like stellar evolution to self–consistent halo models that include adiabatic contraction and halo depletion; (b) include multi–dimensional effects (rotation, magnetic fields) and time–dependent accretion histories; and (c) produce synthetic spectra and light curves to map the JWST observables corresponding to the inflated and post–DM–fuel phases.

In summary, variations in WIMP mass and scattering cross section, together with the ambient DM density, produce qualitatively distinct protostellar evolutionary channels. When the ambient DM density and capture efficiency exceed the thresholds identified here, dark–matter annihilation can (i) inflate protostars, (ii) suppress early ionizing feedback, (iii) delay GR collapse to much higher masses, and (iv) therefore enable the formation of heavy black hole seeds. These conclusions provide a quantitative and testable framework for evaluating the contribution of Pop III.1 stars to early supermassive black hole assembly and motivate targeted follow–up studies coupling stellar evolution, halo dynamics and radiative signatures.

\begin{acknowledgements}
This work formed the basis of K.T.'s Master's thesis at Gothenburg University (GU). We thank Maria Sundin for being the GU examiner of the thesis. D.N. and J.C.T. acknowledges funding from the Virginia Institute for Theoretical Astrophysics (VITA), supported by the College and Graduate School of Arts and Sciences at the University of Virginia.
D.N. also acknowledges support from a VITA-Origins postdoctoral fellowship. 
J.C.T. also acknowledges support from ERC Advanced Grant MSTAR (788829) and funding from the Virginia Institute for Theoretical Astrophysics (VITA), supported by the College and Graduate School of Arts and Sciences at the University of Virginia.

\end{acknowledgements}

\bibliographystyle{aa}
\bibliography{biblio}

\end{document}